\newif\ifarXiv
\arXivtrue 

\ifarXiv
\documentclass{frontiersENGARXIV} 

\usepackage{url}

\else

\documentclass{frontiersENG} 

\usepackage{url,lineno}
\linenumbers

\fi

\usepackage{color}

\newcommand{\TCI}{\widetilde{CI}}
\newcommand{\TSI}{\widetilde{SI}}
\newcommand{\TUI}{\widetilde{UI}}

\usepackage{setspace} 
\usepackage{nameref}
\newcommand{\eq}[1]{eq.~\ref{eq:#1}}
\newcommand{\secRef}[1]{Section \ref{sec:#1}}
\newcommand{\fig}[1]{Figure ~\ref{fig:#1}}

\usepackage{todo}
\usepackage{graphicx}
\usepackage{subfigure} 

\copyrightyear{}
\pubyear{}

\copyrightyear{}
\pubyear{}

\ifarXiv
\def\journal{AI and Robotics (submitted)}
\else
\def\journal{AI and Robotics}
\fi

\def\DOI{}
\def\articleType{Review}
\def\keyFont{\fontsize{8}{11}\helveticabold }
\def\firstAuthorLast{Wibral {et~al.}} 
\def\Authors{Michael Wibral\,$^{1,*}$, Joseph T. Lizier$^{2}$, Viola Priesemann,$^{3,4}$}
\def\Address{$^{1}$MEG Unit, Brain Imaging Center, Goethe University, Frankfurt am Main, Germany \\
$^{2}$CSIRO Digital Productivity Flagship, Marsfield, NSW, Australia\\
$^{3}$Department of Non-linear Dynamics, Max Planck Institute for Dynamics and Self-Organization, G\"{o}ttingen, Germany\\
$^{4}$ Bernstein Center for Computational Neuroscience, G\"{o}ttingen, Germany
}
\def\corrAuthor{Michael Wibral}
\def\corrAddress{MEG Unit, Brain Imaging Center, Goethe University, Heinrich-Hoffmann Strasse 10, Frankfurt am Main, D-602528, Germany}
\def\corrEmail{wibral@em.uni-frankfurt.de}

\begin{document}
\onecolumn
\firstpage{1}

\title[Bits from Biology]{Bits from Biology for Computational Intelligence}
\author[\firstAuthorLast ]{\Authors}
\address{}
\correspondance{}
\extraAuth{}
\topic{}

\maketitle

\begin{abstract}
Computational intelligence is broadly defined as biologically-inspired computing. Usually, inspiration is drawn from neural systems. This article shows how to analyze neural systems using information theory to obtain constraints that help identify the algorithms run by such systems and the information they represent. Algorithms and representations identified information-theoretically may then guide the design of biologically inspired computing systems (BICS). The material covered includes the necessary introduction to information theory and the estimation of information theoretic quantities from neural data. We then show how to analyze the information encoded in a system about its environment, and also discuss recent methodological developments on the question of how much information each agent carries about the environment either uniquely, or redundantly or synergistically together with others. Last, we introduce the framework of local information dynamics, where information processing is decomposed into component processes of information storage, transfer, and modification -- locally in space and time. We close by discussing example applications of these measures to neural data and other complex systems.

\tiny
 \keyFont{ \section{Keywords:} Information Theory, Local Information Dynamics, Partial Information Decomposition, Neural Systems, Computational Intelligence, Biologically Inspired Computing, Artificial Neural Networks} 
\end{abstract}

\section{Introduction}
Computational intelligence (CI) is broadly defined as biologically-inspired computing. CI often must deal with ill-posed problems and the field of CI draws inspiration from natural information processing systems -- as these cannot afford the luxury to dismiss any problem that happens to cross their path as `ill-posed'. Instead, natural systems have evolved algorithms to approximately solve problems relevant to them: algorithms that are adapted to their often limited resources and that yield `good enough' solutions. These algorithms may then serve as an inspiration for artificial information processing systems to solve similar problems under tight constraints of computational power, data availability, and time.

One way to use this inspiration is to copy and incorporate as much biological detail as possible in the artificial system, in the hope to also copy the emergent information processing of the biological system.  However, already small errors in copying the parameters of a system may compromise success. Therefore, it may be useful to derive inspiration also in a more abstract way, that is directly linked to the information processing carried out by a biological system. But how can can we gain insight into this information processing without caring for its biological implementation?

The formal language to quantitatively describe and dissect information processing -- in any system -- is provided by information theory. For our particular question we can exploit the fact that information theory does not care about the nature of variables that enter the computation or information processing. Thus, it is in principle possible to treat all relevant aspects of biological computation, and of biologically inspired computing systems, in one natural framework.

Here, we will first review some information theoretic preliminaries. Then we will systematically present how to analyze biological computing systems, especially neural systems, using methods from information theory and discuss how these information theoretic results can inspire artificial computing systems. We will close by a brief review of studies where this information theoretic point of view has served this goal.

\section{Information Theory in Neuroscience}
\subsection{Information Theoretic Preliminaries}
In this section, we introduce the necessary terminology, and notation, and define basic information theoretic quantities that later analyses build on. Experts in information theory may proceed immediately to 
\secRef{UseInfoNeuro} 
which discusses the use of information theory in neuroscience.

\subsubsection{Terminology and Notation} 
To analyze neural systems and biologically-inspired computing systems (\emph{BICS}) alike, and to show how the analysis of one can inspire the design of the other, we have to establish a common terminology. Neural systems and BICS have the common property that they are composed of various smaller parts that interact. These parts will be called  \emph{agents} in general, but we will also refer to them as \emph{neurons} or \emph{brain areas} where appropriate.  The collection of all agents will be referred to as the \emph{system}.

We define that an agent $\mathcal{X}$ in a system produces an observed time series $\{x_1, \ldots,x_t,\ldots, x_N\}$ which is sampled at time intervals $\Delta$. For simplicity we choose $\Delta=1$, and index our measurements by $t \in \{1...N\}\subseteq\mathbb{N}$. The time series is understood as a realization of a \emph{random process} $\texttt{X}$. The random processes is a collection of random variables (RVs) $X_t$, sorted by an integer index ($t$). Each RV $X_t$, at a specific time $t$, is described by the set of all its $J$ possible outcomes $\mathcal{A}_{X_t}=\{a_1, \ldots a_j \ldots  a_J\}$, and their associated probabilities $p_{X_t}(x_t=a_j)$. Since the probabilities of an outcome $p_{X_t}(x_t=a_j)$ may change with $t$ in nonstationary random processes,
we indicate the RV the probabilities belong to by subscript: $p_{X_t}(\cdot)$. In sum, the physical agent $\mathcal{X}$ is conceptualized as a random process $\texttt{X}$, composed of a collection of RVs $X_t$, that produce realizations $x_t$, according to the probability distributions $p_{X_t}(x_t)$. When referring to more than one agent, the notation is generalized to  $\mathcal{X}, \mathcal{Y}, \mathcal{Z}, \ldots$~.
An overview of the complete notation can be found in table \ref{tab:notation}.

\begin{table}[!t]
\textbf{\refstepcounter{table}\label{tab:notation} Table \arabic{table}.}{ Notation}

\processtable{ }
{\begin{tabular}{ll}\toprule
 $\mathcal{X, Y, Z}$ & agent in a system\\
\texttt{X, Y, Z} & random process \\
$X, Y, Z$ or $X_t, Y_t, Z_t$ & random variable (at time point $t$) \\
&	Whenever necessary, the index $t$ is detailed as $t_1, t_2, ..., t_k$. \\
 & For stationary processes, the index $t$ can be omitted.\\
$x, y, z$ or $x_t, y_t, z_t$ & realization of the random variable (at time point $t$) \\
$a_j$ 	& specific outcome of a random variable $x$\\
$p_{X_t}(x_t=a_j)$ & probability that $X_t$ has a specific outcome $a_j$\\
$\mathcal{A}_{X_t} = \{ a_1, \ldots a_j, \ldots a_J\} $ & set of all possible outcomes of $X_t$ \\
$\texttt{X}^{(c)}, X_t^{(c)}$ & cyclostationary process and cyclostationary random variable \\
$\texttt{X}^{(s)}, X^{(s)}$ & stationary process and stationary random variable \\
$\textbf{X}_t, \textbf{x}_t$ & state space representation of \texttt{X} at $t$\\	
$\textbf{X}^-_{t-u}$ & state space representation of \texttt{X} at $t-u$;\\
~& the superscript minusH serves as a reminder\\
& that $\textbf{X}^-_{t-u}$ is in the past of $X_t$\\
$u$ & assumed interaction delay between two processes  \\		
$\delta$ & physical or true interaction delay between two processes \\
$S_i$, $R_j$ & random variables referring to stimuli ($S_i$) or responses ($R_j$) \\
$\textbf{R} = \{R_1,R_2\} $ & joint variable (in this example of two responses) \\	
$H(X)$ & entropy \\
$H(X|Y)$ & conditional entropy \\
$h(x)$ & information content \\				
$h(x|y)$ & conditional information content \\
$I(X:Y)$ & mutual information \\
$I(X:Y|Z)$ & conditional mutual information \\
~ & note that the colon is used to separate the random variables between which we compute $I$\\
$i(x:y)$ & local mutual information \\		
$i(x:y|z)$ & local conditional mutual information \\
$X,Y$ & the comma is used to separate random variables \\
$X_1,X_2; Y_1, Y_2$ & the semicolon is used to separate sets of random variables \\

\end{tabular}}{}
\end{table}

\subsubsection{Estimation of Probability Distributions for Stationary and Non-stationary Random Processes}
\label{sec:EstProb}
In general, the probability distributions of the $X_t$ are unknown. Since knowledge of these probability distributions is essential to computing any information theoretic measure, the probability distributions have to be estimated from the observed realizations of the RVs, $x_{t}$. This is only possible if we have some form of replication of the processes we wish to analyze. From such replications the probabilities are estimated for example by counting relative frequencies, or by density estimation \citep{koz87.1,Kraskov2004,victor02.1}.

In general, the probability $p_{X_t}(x_t=a_j)$ to obtain the j-th outcome $x_{t}=a_j$ at time $t$, has to be estimated from replications of the processes at the same time point $t$, i.e. via an ensemble of \emph{physical} replications of the systems in question. These replications can be often be obtained in BICS via multiple simulation runs or even physical replications if the systems in question are very small and/or simple. For complex physically embodied BICS and neural systems, generating a sufficient number of replications of a process is often impossible. Therefore, one either resorts to \emph{repetitions} of parts of the process in time, to the generation of cyclostationary processes, or even assumes stationarity. All three possibilities will be discussed in the following.

\paragraph{General Repetitions in Time.} If our random process can be repeated in time, then the probability to obtain the value $x_{t}=a_j$ can be estimated from observations made at a sufficiently large set $\mathcal{M}$ of time points $t+k$, where we know by design of the experiment that the process repeated itself. That is, we know that RVs $X_{t+k}$ at certain time points $t+k$ have probability distributions identical to the distribution at $t$ that is of interest to us:

\begin{equation}
 \forall \ t ~ \exists \mathcal{M} \subseteq \mathbb{N} \wedge \mathcal{M}\neq\varnothing : p_{X_t}(a_j)=p_{X_{t+k}}(a_j)\quad \forall \ k \in \mathcal{M}, a_j \in \mathcal{A}_{X_t}.
\end{equation}

\noindent If the set $\mathcal{M}$ of times $t_k$ that the process is repeated at is large enough, we obtain a reliable estimate of $p_{X_t}(\cdot)$.

\paragraph{The Cyclostationary Case.} Cyclostationarity can be understood as a specific form of repeating parts of the random process, where the repetitions occur after regular intervals $T$. For cyclostationary processes \texttt{X}$^{(c)}$ we assume \citep{Gardner1994,Gardner2006}, that there are RVs $X^{(c)}_{t+nT}$ at times $t+nT$ that have the same probability distribution as $X^{(c)}_{t}$:
\begin{equation}
\label{eg:cyclo}
 \exists T \in \mathbb{N}  : p_{X_t}(a_j)=p_{X_{t+nT}}(a_j)\quad \forall t,n \in  \mathbb{N},~ t<T, ~a_j \in \mathcal{A}_{X_t}.
\end{equation}
This condition guarantees that we can estimate the necessary probability distributions $p_{X_t}(\cdot)$ of the RV $X^{(c)}_t$ by looking at other RVs $X^{(c)}_{t+nT}$ of the process \texttt{X}$^{(c)}$.

\paragraph{Stationary Processes.} Finally, for stationary processes \texttt{X}$^{(s)}$, we can substitute $T$  in eq. \ref{eg:cyclo} by $T=1$ and:
\begin{equation}
\label{eg:statio}
 p_{X_t}(a_j)=p_{X_{t+n}}(a_j)\quad \forall t, n \in  \mathbb{N} , a_j \in \mathcal{A}_{X_t}.
\end{equation}

\noindent In the stationary case the probability distribution $p_{X_t}(\cdot)$ can be estimated from the entire set of measured realizations $x_t$.  Thus, we will drop the subscript index indicating the specific RV, i.e. $p_{X_t}(\cdot)=p(\cdot)$, $X_t=X$ and $x_t=x$ when the process is stationary, and also when stationarity is irrelevant (e.g. when talking only about a single RV).

\subsubsection{Basic Information Theory}
\label{sec:basicInfoTheory}
Based on the above definitions we now define the necessary basic information theoretic quantities. To put a focus on the often neglected \emph{local} information theoretic quantities that will become important later on, we will start with the Shannon information content of a realization of a RV.

To this end, we assume a (potentially nonstationary) random process \texttt{X} consisting of $X_1,X_2,\ldots,X_N$.  The law of total probaility states that
\begin{equation}
 \sum_{x_1,x_2,\ldots,x_N} p(x_1,x_2,\ldots,x_N)=1~,
\end{equation}
\noindent and the product rule yields
\begin{align}
 \sum_{x_1} p(x_1)\sum_{x_2,\ldots,x_N}p(x_2,\ldots,x_N |x_1)=1\\
 \intertext{with}
 \sum_{x_2,\ldots,x_N}p(x_2,\ldots,x_N |x_1)=1~.
\end{align}
\noindent All realizations of the process starting with a specific $x_1$ thus together have probability mass
\begin{equation}
p(x_1)\sum_{x_2,\ldots,x_N}p(x_2,\ldots,x_N |x_1)=p(x_1)~,
\end{equation}
\noindent and occupy a fraction of $p(x_1)/1$ in the original probability space. Obtaining $x_1$ can therefore be interpreted as informing us that the the full realization lies in this fraction of the space. Thus, the reduction in uncertainty, or the information gained from $x_1$ must be a function of $1/p(x_1)$. To ensure that subsequent realizations from indepenent RVs yield additive amounts of information, we take the logarithm of this ratio to obtain the \emph{Shannon information content} \citep{Shannon1948} (also see \citet{mackay2003Book}) which measures the information provided by a single realization $x_i$ of a RV $X_i$:

\begin{equation}
\label{eq:Shannon}
 h(x_i)=\log \frac{1}{p(x_i)}.
\end{equation}
Typically, we take $\log_2$ giving units in \emph{bits}.

The \emph{average} information content of a RV $X_i$ is called the \emph{entropy} $H$:
\begin{equation}
 H(X_i)=\sum_{x_i \in \mathcal{A}_{x_i}} p(x_i) \log \frac{1}{p(x_i)}.
\end{equation}

The information content of a specific realization $x$ of $X$, given we already know the outcome $y$ of another variable $Y$, which is not necessarily independent of $X$, is called \emph{conditional information content}:
\begin{equation}
 h(x|y)=\log \frac{1}{p(x|y)}
\end{equation}

Averaging this for all possible outcomes of $X$, given their probabilities $p(x|y)$ after the outcome $y$ was observed and averaging then over all possible outcomes $y$, that occur with $p(y)$, yields the \emph{conditional entropy}:
\begin{equation}
 H(X|Y)=\sum_{y \in \mathcal{A}_Y} p(y) \sum_{x \in \mathcal{A}_X} p(x|y) \log \frac{1}{p(x|y)}=\sum_{x \in \mathcal{A}_X,y \in \mathcal{A}_Y}p(x,y) \log \frac{1}{p(x|y)}
\end{equation}

The conditional entropy $H(X|Y)$ can be described from various perspectives: $H(X|Y)$ is the average amount of information that we get from making an observation of $X$ after having already made an observation of $Y$. In terms of uncertainties $H(X|Y)$ is the average remaining uncertainty in $X$ once $Y$ was observed. We can also say $H(X|Y)$ is the information in $X$ that can not be directly obtained from $Y$.

The conditional entropy can be used to derive the amount of information directly \emph{shared} between the two variables $X,Y$. This is because the mutual information of two variables $X$, $Y$, $I(X:Y)$, is the the total average information in one variable ($H(X)$) minus the average information in this variable that can not be obtained from the other variable ($H(X|Y)$). Hence the \emph{mutual information} (MI) is defined as:
\begin{equation}
 I(X:Y)=H(X)-H(X|Y)=H(Y)-H(Y|X)
\end{equation}
\noindent Similarly to conditional entropy we can also define a \emph{conditional mutual information} between two variables $X,Y$, given the value of a third variable $Z$ is known:
\begin{equation}
\label{eq:mi}
 I(X:Y | Z)=H(X|Z)-H(X|Y,Z)
\end{equation}

The above measures of mutual information are averages. Although average values
are used more often than their localized counterparts, it is perfectly valid to inspect local values for MI (like the information content $h$, above). This `localizability' was in fact a requirement that both Shannon and Fano postulated for proper information theoretic measures \citep{Fano1961book,Shannon1948}, and there is a growing trend in neuroscience \citep{Lizier2011a} and in the theory of distributed computation \citep{LizierBook,Lizier2014} to return to local values. For the above measures of mutual information the localized forms are listed in the following.

The  \emph{local} mutual information $i(x:y)$ is defined as:
\begin{equation}
\label{eq:lmi}
 i(x:y)=\log \frac{p(x,y)}{p(x)p(y)}=\log \frac{p(x|y)}{p(x)}
\end{equation}
while the \emph{local} conditional mutual information is defined as:
\begin{equation}
\label{eq:lcmi}
 i(x:y | z)=\log \frac{p(x|y, z)}{p(x|z)}
\end{equation}

\noindent When we take the expected values of these local measures, we obtain mutual and conditional mutual information. These measures are called local, because they allow one to quantify mutual and conditional mutual information between \emph{single realizations}. Note, however, that the probabilities $p(\cdot)$ involved in equations \ref{eq:lmi} and \ref{eq:lcmi} are \emph{global} in the sense that they are representative of all possible outcomes. In other words, a valid probability distribution has to be estimated irrespective of whether we are interested in average or local information measures.
We also note that local MI and local conditional MI may be negative, unlike their averaged forms \citep{Fano1961book,Lizier2014}. This occurs for the local MI where the measurement of one variable is \textit{misinformative} about the other variable, i.e. where the realization $y$ \textit{lowers} the probability $p(x|y)$ below the initial probability $p(x)$. This means that the observer expected $x$ less after observing $y$ than before, but $x$ occurred nevertheless. Therefore, $y$ was misinformative about $x$.

\subsubsection{Estimating information theoretic quantities from data}
Before we advance to specific information theoretic analyses of neural data, it must be stressed that the estimation of information theoretic measures from finite data is a difficult task. The naive estimation of probabilities by empirically observed frequencies, followed by plugging of these probabilities into the above definitions almost inevitably leads to serious bias problems \citep{Treves1995,Panzeri2007,victor02.1}. This situation can be improved to some degree by using binless density estimators \citep{victor02.1,koz87.1,Kraskov2004}. However, ususally statistical testing against surrogate data or empirical control data will be necessary to judge whether a non-zero value of a measure indicates an effect or just the bias (see e.g.\,\citet{Lindner2011}).

\subsubsection{Signal Representation and State Space Reconstruction}
\label{sec:states}
The random processes that we analyze in the agents of a computing system usually have memory. This means that the RVs that form the process are no longer independent, but depend on variables in the past. In this setting, a proper description of the process requires to look at the present and past RVs jointly. In general, if there is any dependence between the $X_t$, we have to form the smallest collection of variables $\mathbf{X_t}=(X_{t},X_{t_1},X_{t_2},\ldots,X_{t_i},\ldots )$ with $t_i<t$ that jointly make $X_{t+1}$ conditionally independent of all $X_{t_k}$ with $t_k<\min(t_i)$, i.e.:
\begin{equation}
\label{eq:embedding1}
\begin{split}
p(x_{t+1},x_{{t_k}} | \mathbf{x_{t}})= p(x_{t+1} | \mathbf{x_{t}}) p(x_{{t_k}} | \mathbf{x_{t}}), \\
 \textrm{i.e.\ \ \ } p(x_{t+1} | x_{{t_k}}, \mathbf{x_{t}})= p(x_{t+1} | \mathbf{x_{t}}) \\
 \forall t_k < \min(t_i),~\forall x_{t+1}\in\mathcal{A}_{X_{t+1}},~\forall x_{{t_k}}\in\mathcal{A}_{X_{t_k}},~\forall \mathbf{x_{t}} \in \mathcal{A}_{\mathbf{x_{t}}}
\end{split}
\end{equation}
 \noindent A realization $\mathbf{x_{t}}$ of such a sufficient collection $\mathbf{X_{t}}$  of past variables is called a \emph{state} of the random process \texttt{X} at time $t$.
 
 A sufficient collection of past variables, also called a delay embedding vector, can always be reconstructed from scalar observations for low dimensional deterministic systems, as shown by \citet{Takens1981}. Unfortunately, most real world systems have high-dimensional dynamics rather than being low-dimensional deterministic. For these systems it is not obvious that a delay embedding similar to Taken's approach would yield the desired results. In fact, many systems require an infinite number of past random  variables when only a scalar observable of the high-dimensional stochastic process is accessible \citep{Ragwitz2002}. Nevertheless, the behavior of scalar observables of most of these systems can be approximated well by a finite collection of such past variables for all practical purposes \citep{Ragwitz2002}; in other words, these systems can be approximated well by a finite order, one-dimensional Markov-process according to \eq{embedding1}. 
 
Note that without proper state space reconstruction information theoretic analyses will almost inevitably miscount information in the random process. Indeed, the importance of state space reconstruction cannot be overstated; for example, a failure to reconstruct states properly lead to false positive findings and reversed directions of information transfer as shown in \citep{VWL+11}; imperfect state space reconstruction is also the cause of failure of transfer entropy analysis demonstrated in \citep{Smirnov2013}; and has been shown to impede the otherwise clear identification of coherent moving structures in cellular automata as information transfer entities \citep{Lizier2008}.

In the remainder of the text we therefore assume proper state space reconstruction. The resulting state space representations are indicated by bold case letters, i.e. $\mathbf{X_t}$ and $\mathbf{x_{t}}$ refer to the state variables of \texttt{X}.

\subsection{Why Information Theory in Neuroscience?}
\label{sec:UseInfoNeuro}
It is useful to organize our understanding of neural (and biologically-inspired) computing systems into three major levels, originally proposed by David Marr \citep{MarrBook}, and to then see at which level information theory provides insights:

\begin{itemize}
 \item At the level of the task the neural system or the BICS is trying to solve (\emph{task} level\footnote{Called the \emph{``computational level''} by Marr originally. This terminology, however, collides with other meanings of computation used in this text. }) we ask what information processing problem a neural system (or a part of it) tries to solve. Such problems could for example be the detection of edges or objects in a visual scene, or maintaining information about an object after the object is no longer in the visual scene. It is important to note that questions at the task level typically revolve around entities that have a direct meaning to us, e.g. objects or specific object properties used as stimulus categories, or operationally defined states, or concepts such as attention or working memory. An example of an analysis carried out purely at this level is the investigation of whether a person behaves as an optimal Bayesian observer (see references in \citet{Knill2004}).

\item At the \emph{algorithmic} level we ask what entities or quantities of the task level are represented by the neural system and how the system operates on these representations using algorithms. 
For example, a neural system may represent either absolute luminance or changes of luminance of the visual input. An algorithm operating on either of these representations may for example then try to identify an object in the input that is causing the luminance pattern either by a brute force comparison to all luminance patterns ever seen (and stored by the neural system). Alternatively, it may try to further transform the luminance representation via filtering etc.\,, before inferring the object via a few targeted comparisons.

\item  At the (biophysical) \emph{implementation} level, we ask how the representations and algorithms are implemented in neural systems. Descriptions at this level are given in terms of the relationship between various biophysical properties of the neural system or its components, e.g. membrane currents or voltages, the morphology of neurons, spike rates, chemical gradients etc.\,. A typical study at this level might aim for example at reproducing observed physical behavior of neural circuits, such as gamma-frequency ($>$40~Hz) oscillations in local field potentials by modeling the biophysical details of these circuits from ground up \citep{Markram2006}. 
\end{itemize}

This separation of levels of understanding served to resolve important debates in neuroscience, but there is also growing awareness of a specific shortcoming of this classic view: results obtained by careful study at any of these levels do not constrain the possibilities at any other level (see the after-word by Poggio in \citet{MarrBook}). For example, the task of winning a game of Tic-Tac-Toe (task level) can be reached by a brute force strategy (algorithmic level) that may be realized in a mechanical computer (implementation level) \citep{dewdney1989tinkertoy}. Alternatively, the very same task can be solved by flexible rule use (algorithmic level) realized in biological brains (implementation level) of young children \citep{Crowley1993flexible}.

As we will see, missing relationships between Marr's levels can be filled in by information theory: In 
\secRef{NeuralCode} 
we show how to link the task level and the implementation level by computing various forms of mutual information between variables at these two levels. These mutual informations can be further decomposed into the contributions of each agent in a multi-agent system, as well as information carried  jointly. This will be covered in \secRef{PID}. In \secRef{DistComp} we use local information measures to link neural activity at the implementation level to components of information processing at the algorithmic level, such as information storage, and transfer. This will be done \emph{per agent and time step} and thereby yields a sort of information theoretic ``footprint'' of the algorithm in space and time. To be clear, such an analyses will only yield this ``footprint''-- not identify the algorithm itself. Nevertheless, this footprint is a useful constraint when identifying algorithms in neural systems, because various possible algorithms to solve a problem will clearly differ with respect to this footprint. \secRef{InfoMod} covers current attempts to define the concept of information modification. We close by a short review of some example applications of information theoretic analyses of neural data, and describe how they relate to Marr's levels.

\section{Analyzing Neural Coding}
\label{sec:NeuralCode}
\subsection{Neural Codes for External Stimuli}
As introduced above, information theory can serve to bridge the gap between the task level, where we deal with properties of a stimulus or task that bear a direct meaning to us, and the implementation level, where we recorded physical indices of neural activity, such as action potentials. To this end we use mutual information (eq. \ref{eq:mi}) and derivatives thereof to answer questions about neural systems like these:

\begin{enumerate}
 \item Which (features of) neural responses ($R$) carry information about which (features of) stimuli ($S$)?
 \item How much does an observer of a specific neural response \emph{r}, i.e. a receiving brain area, change its beliefs about the identity of a stimulus $s$, from the initial belief $p(s)$ to the posterior belief $p(s|r)$ after receiving the neural response $r$?
 \item Which specific neural response \emph{r} is particularly informative about an unknown stimulus $s$ from a certain set of stimuli?
 \item Which stimulus $s$ leads to responses that are informative about this very stimulus, i.e. to responses that can ``transmit'' the identity of the stimulus to downstream neurons?
\end{enumerate}

\noindent The empirical answers to these questions  bear important implications for the design of BICS. For example, the encoding of an enviroment in a BICS maybe modeled on that of a neural system that successfully lives in the same environment.  In the following paragraphs we will show how to answer the above questions 1-4 using information theory.

\paragraph*{1. Which neural responses (R) carry information about which stimuli (S)?} 
This question can be easily answered by computing the mutual information $I(S:R)$ between stimulus identity and neural responses. Despite its deceptive simplicity, computing this mutual information can be very informative about neural codes. This is because both the description of what constitutes a stimulus and a response rely on what we consider to be their relevant features. For example, presenting pictures of fruit as stimulus set, we could compute the mutual information between neural responses and the stimuli described as red versus green fruit or described as apples versus pears. The resulting mutual information will differ between these two descriptions of the stimulus set -- allowing us to see how the neural system partitions the stimuli. Likewise, we could extract features $F_i(r)$ of neural responses $r$, such as the time of the first spike (e.g. \citep{Johansson2004}), or the relative spike times \citep{Havenith2011,OKeefe1993}). Comparing the mutual information for two features $I(S:F_1(R)),I(S:F_2(R))$ allows to identify the feature carrying most information. This feature potentially is the one also read out internally by other stages of the neural system. However, when investigating individual stimulus or response features, one should also keep in mind that several stimulus or response features might have to be considered jointly as they could carry synergistic information (see \secRef{PID}, below).

\paragraph*{2. How much does an observer of a specific neural response $r$, i.e. a receiving neuron or brain area, change its beliefs about the identity of a stimulus $s$, from the prior belief $p(s)$ to the posterior belief $p(s|r)$ after receiving the neural response $r$?} This question is natural to ask in the setting of Bayesian brain theories \citep{Knill2004}. Since this question addresses a quantity associated with a specific response ($r$), we have to decompose the overall mutual information  between the stimulus variable and the response variable ($I(S:R)$) into more specific information terms. As this question is about a difference in probability distributions, before and after receiving $r$, it is naturally expressed in terms of a Kullback-Leibler divergence between  $p(s)$ and  $p(s|r)$. The resulting measure is called the \emph{specific surprise} $i_{sp}$ \citep{DeWeese1999}:
 \begin{equation}
  \label{eq:suprise}
  i_{sp}(S:r)=\sum_{s \in \mathcal{A}_s}p(s|r)\log\frac{p(s|r)}{p(s)}~.
 \end{equation}
\noindent It can be easily verified that $I(S:R)=\sum_r p(r)i_{sp}(S:r)$. Hence $i_{sp}$ is a valid decomposition of the mutual information into more specific, response dependent contributions.
Similarly, we have $i_{sp}(S:r)=\sum_{s}p(s|r) i(s:r)$, giving the relationship between the (fully) localized MI (\eq{lmi}) and $i_{sp}(S:r)$ as a \emph{partially}-localized MI.
As a Kullback- Leibler divergence, $i_{sp}$ is always positive or zero:
\begin{equation}
 i_{sp}(S:r)\geq 0
\end{equation}
\noindent This simply indicates that any incoming response will either update our beliefs (leading to a positive Kullback-Leibler divergence), or not (in which case the Kullback-Leibler divergence will be zero). From this it immediately follows that $i_{sp}$ cannot be additive: if of two subsequent responses $r_1$, $r_2$, the first leads us to update our beliefs about $s$ from $p(s)$ to $p(s|r_1)$, but the second leads us to revert this update, i.e. $p(s|r_1,r_2)=p(s)$ then $i_{sp}(S:r_1,r_2)=0\neq i_{sp}(S:r_1)+i_{sp}(S:r_2|r_1)$. Loosely speaking, a series of surprises and belief updates does not necessarily lead to a better estimate. This fact has been largely overlooked in early applications of this measure in neuroscience as pointed out by \citet{DeWeese1999}. Some caution is therefore necessary when interpreting results from the literature before 1999 that were obtained using this particular decomposition of the mutual information.

\paragraph*{3. Which specific neural response $r$ is particularly informative about an unknown stimulus from a certain set of stimuli?} This question asks how much the knowledge about $r$ is worth in terms of an \emph{uncertainty reduction} about $s$, i.e. an information gain. In contrast to the question about an update of our beliefs above, we here ask whether this update increases or reduces uncertainty about $s$. This question is naturally expressed in terms of conditional entropies, comparing our uncertainty before the response, $H(S)$, with our uncertainty after receiving the specific response $r$, $H(S|r)$. The resulting difference is called the (response-) \emph{specific information} $i_{r}(S:r)$ \citep{DeWeese1999}:
\begin{equation}
 i_r(S:r)=H(S)-H(S|r),
\end{equation}
\noindent where $H(S|r)=\sum_{s} p(s|r) \log\frac{1}{p(s|r)}$. Again it is easily verified that $I(S:R)=\sum_r p(r)i_{r}(S:r)$. 
However, here the individual contributions, $i_r(S:r)$, are not necessarily positive. This is because a response $r$ can lead from a probability distribution $p(s)$ with a low entropy $H(S)$ to some $p(s|r)$ with a high entropy $H(S|r)$. Accepting such `negative information' terms makes the measure additive for two subsequent responses:
\begin{equation}
 i_{r}(S:r_1,r_2)=i_{r}(S:r_1)+i_{r}(S:r_2|r_1)~.
\end{equation}
\noindent The negative contributions $i_{r}(S:r)$ can be interpreted as responses $r$ that are mis-informative in the sense of an increase in uncertainty about the \emph{average} outcome of $S$ (compare the misinformation on the fully local scale indicated by negative $i(x:y)$; see \secRef{basicInfoTheory}).

 \paragraph*{4. Which stimulus $s$ leads to responses $r$ that are informative about the stimulus itself?} In other words, which stimulus is reliably associated to responses that are relatively unique for this stimulus, so that we know about the occurrence of this specific stimulus from the response unambiguously. Here we ask about stimuli that are being encoded well by the system, in the sense that they lead to responses that are informative to a downstream observer. In this type of question a response is considered informative if it strongly reduces the uncertainty about the stimulus, i.e. if it has a large $i_{r}(S:r)$. We then ask how informative the responses for a given stimulus $s$ are on average over all responses that the stimulus elicits with probabilities $p(r|s)$:
 \begin{equation}
  i_{SSI}(s:R)=\sum_{r \in \mathcal{A}_r} ~ p(r|s)i_{r}(S:r)~.
 \end{equation}
The resulting measure $i_{SSI}(s:R)$ is called the \emph{stimulus specific information (SSI)} \citep{Butts2003}. Again it can be verified easily that $I(S:R)=\sum_s p(s)i_{SSI}(s:R)$, meaning that $i_{SSI}$ is another valid decomposition of the mutual information. Just as the response specific information terms that it is composed of, the stimulus specific information can be negative \citep{Butts2003}.

\begin{figure}[bh]
\begin{center}
\includegraphics[width=10cm]{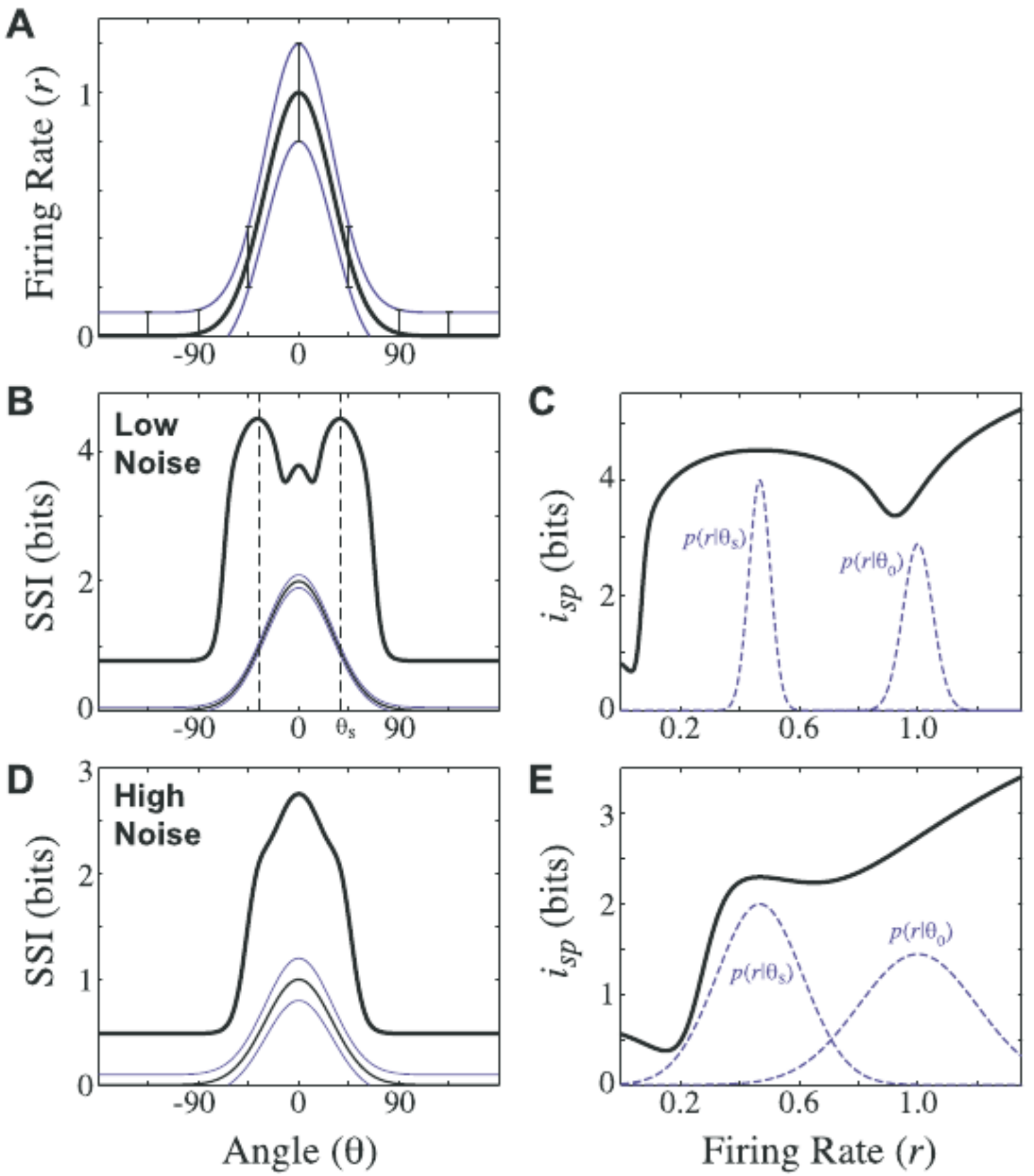}
\end{center}
\textbf{\refstepcounter{figure}\label{fig:Butts} Figure \arabic{figure}.}{ Stimulus specific surprise ($i_{sp}$) and stimulus specific information ($i_{SSI}$) of an orientation tuned model neuron under two different noise regimes. (A) Tuning curve: mean firing rate (thick line), standard deviation (thin lines) versus stimulus orientation ($\Theta$). Repeated in for (B,D) for clarity. (B) The stimulus specific information $i_{SSI}$ (indicated as SSI) is maximal in regions of high slope of the tuning curve for the low noise case; (D) for the high noise case  $i_{SSI}$ (indicated as SSI) is maximal at the peak of the  tuning curve. (C,E) The corresponding values of the stimulus specific surprise $i_{sp}$ and and the relevant conditional probability distributions. Figure reproduced from \citep{Butts2006}. Creative Commons (CC BY) Attribution License.}
\end{figure}

The stimulus specific information has been used to investigate which stimuli are encoded well in neurons with a specific tuning curve; it was demonstrated that the specific stimuli that were encoded best changed with the noise level of the responses \citep{Butts2006} (Figure \ref{fig:Butts}). Results of this kind may for example be important to consider in the design of BICS that will be confronted with varying levels of noise in their environments.

\hspace{1cm}
\newpage

\subsection{Importance of the stimulus set and response features}
It may not immediately be visible in the above equations, but central quantities of the above treatment, such as $H(S)$, $H(S|r)$ depend strongly on the choice of the stimulus set $\mathcal{A}_S$. For example, if one chooses to study the human visual systems with a set of ``visual'' stimuli in the far infrared end of the spectrum, $I(S:R)$ will most likely be very small and analysis futile (although done properly, a zero value of $i_{SSI}(s:R)$ for all stimuli will correctly point out that the human visual system does not care or code for any of the infrared stimuli). Hence, characterizing a neural code properly hinges to a large extent on an appropriate choice of stimuli. In this respect, it is safe to assume that a move from artificial stimuli (such as gratings in visual neuroscience) to more natural ones will alter our view of neural codes in the future. A similar argument holds for the response features that are selected for analysis. If any feature is dropped or not measured at all this may distort the information measures above. This may even happen, if the dropped feature, say the exact spike time variable $R_{ST}$, seems to carry no mutual information with the stimulus variable when considered alone, i.e. $I(S:R_{ST})=0$. This is because there may still be synergistic information that can only be recovered by looking at other response variables jointly with $R_{ST}$. For example, it would be possible in principle that neither spike time $R_{ST}$ nor spike rate $R_{SR}$ carry mutual information with the stimulus variable when considered individually, i.e. $I(S:R_{ST})=I(S:R_{SR})=0$. Still, when considered jointly they may be informative: $I(S:R_{ST},R_{SR})>0$. The problem of omitted response features is almost inevitable in neuroscience, as the full sampling of all parts of a neural systems is typically impossible, and we have to work with sub-sampled data. Considering only a subset of (response) variables may dramatically alter the apparent dependency structure in the neural system (see \citet{Priesemann2009} for an example). Therefore, the effects of subsampling should always be kept in mind when interpreting results of studies on neural coding.

\section{Information in Ensemble Coding -- Partial Information Decomposition}
\label{sec:PID}
In neural systems information is often encoded by \emph{ensembles} of agents -- as evidenced by the success of various 'brain reading' and decoding techniques applied to \emph{multivariate} neural data (e.g.\citep{Kriegeskorte2008}). Knowing how this information in the ensemble is distributed over the agents can inform the designer of BICS about strategies to distribute the relevant information about a problem over the available agents. These strategies determine properties like the coding capacity of the system as well as its reliability. For example, a reliable strategy would represent the same information in multiple agents, making their information redundant. In contrast, maximizing capacity would require taking into account the full combinatorial possibilities of states of agents, making their coding synergistic.

Here, we investigate the most basic ensemble of just two agents to introduce the concepts of redundant, synergistic and unique information \citep{PaulWilliamsPID,Bertschinger2014,Harder2013,Griffith2014a,Lizier2013InfoMod}, and note that encoding in larger ensembles is still a field of active research. More specifically, we consider an ensemble of two neurons and their responses, $\{R_1, R_2\}$,  after stimulation with stimuli $s \in \mathcal{A}_S=\{s_1,s_2,\ldots\}$, and try to answer the following questions: 
\begin{enumerate}
 \item What information does $R_i$ provide about $S$? This is the mutual information $I(S:R_i)$ between the responses of one neuron $i$ and the stimulus set.
 \item What information does the joint variable $\mathbf{R}=\{R_1, R_2\}$ provide about $S$? This is the mutual information ~$I(S:R_1,R_2)$ between the joint responses of the two neurons and the stimulus set.
 \item What information does the joint variable $\mathbf{R}=\{R_1, R_2\}$ have about $S$ that we cannot get from observing both variables $R_1$, $R_2$ separately? This information is called the \emph{synergy}, or \emph{complementary information}, of $\{R_1, R_2\}$ with respect to $S$: ~$CI(S : R_1 ; R_2)$.
 \item What information does one of the variables, say $R_1$, hold individually about $S$ that we can not obtain from any other variable ($R_2$ in our case)? This information is the \emph{unique information} of $R_1$ about $S$: ~$UI(S:R_1 \setminus R_2)$.
 \item What information does one of the variables, again say $R_1$, have about S that we could also obtain by looking at  the other variable alone? This information is the \emph{redundant}, or \emph{shared}, information of $R_1$ and $R_2$ about $S$: ~$SI(S: R_1 ; R_2)$. 
\end{enumerate}

Interestingly, only questions 1.\,and 2.\,can be answered using standard tools of information theory such as the mutual information. In fact, the answers to the questions 3.\,to 5.\,, i.e. the quantification of unique, redundant and synergistic information, need new mathematical concepts as will be shown below. 

\begin{figure}[bh]
\begin{center}
\includegraphics[width=12cm]{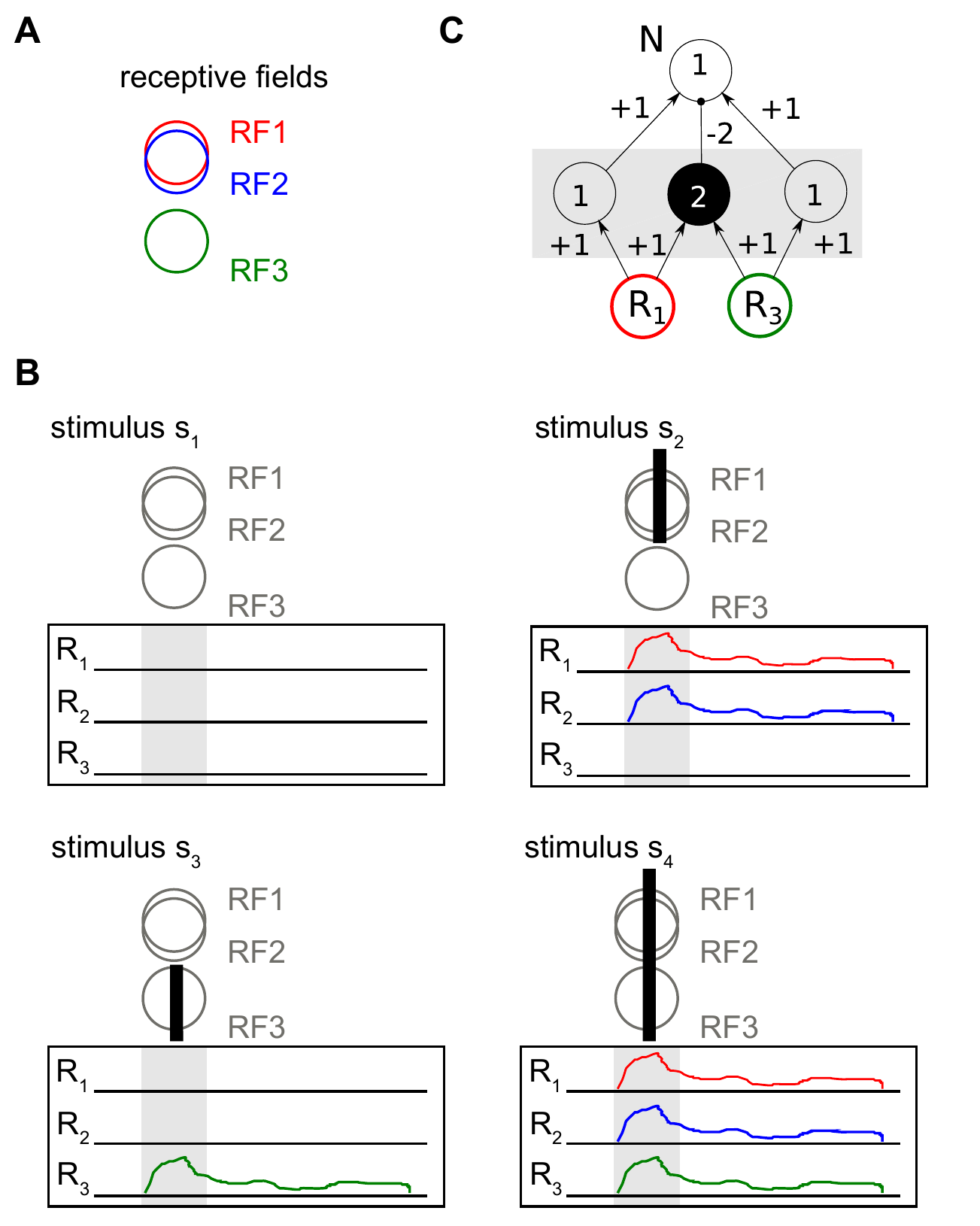}
\end{center}
\textbf{\refstepcounter{figure}\label{fig:PIDneurons} Figure \arabic{figure}.}{ 
Redundant and synergistic neural coding. (A) Receptive fields (RFs) of three neurons $R_1$, $R_2$, $R_3$. (B) Set of four stimuli. (C) Circuit for synergistic coding. Responses of neurons $R_1$, $R_3$ determine the response of neuron $N$ via an XOR-function. In the hidden circuit in between $R_1$, $R_2$ and $N$ open circles denote excitatory neurons, filled circles inhibitory neurons. Numbers in circles are activation thresholds, signed numbers at connecting arrows are synaptic weights.}
\end{figure}

Before we present more details, we would like to illustrate the above questions by a thought experiment where three visual neurons are recorded simultaneously while being stimulated with one of a set of four stimuli (Figure \ref{fig:PIDneurons}). Two of the neurons have almost identical receptive fields ($RF_1$, $RF_2$) while the third one has a collinear but spatially displaced receptive field ($RF_3$) (Figure \ref{fig:PIDneurons} (A)). These neurons are stimulated with one of the following stimuli (Figure \ref{fig:PIDneurons} (B)): $s_1$ does not contain anything at the receptive fields of the three neurons, and the neurons stay inactive; $s_2$ is a small bar with the preferred orientation of neurons 1,2; $s_3$ is a similar small bar, but over the receptive field of neuron 3, instead of 1,2; $s_4$ is a long bar covering all receptive fields in the example.

To make things easy, let us encode responses that we get from these three neurons (colored traces in Figure \ref{fig:PIDneurons} (B)) in binary form, with a ``1'' simply indicating that there was a response in our response window (gray boxes behind the activity traces in Figure \ref{fig:PIDneurons}). If we assume the stimuli to be presented with equal probability ($p(S=s_i)=\frac{1}{4},~i=1,\ldots4$), then the entropy of the stimulus set is $H(S)=2$~(bit). Obviously, none of the information terms above can be larger than these 2 bits. We also see that each neuron shows activity (binary response = 1) in half of the cases, yielding an entropy $H(R_j)=1$ for the responses of each neuron. The responses of the three neurons fully specify the stimulus, and therefore $I(S:{R_1,R_2,R_3})=2$. To see the mutual information between an individual neuron's response and the stimulus we may compute $I(S:R_i)=H(S)-H(S|R_i)$. To do this, we remember $H(S)=2$ and use that the number of equiprobable outcomes for $S$ drops by half after observing a single neuron (e.g. after observing a response $r_1=1$ of neuron 1, two stimuli remain possible sources of this response -- $s_2$ or $s_4$). This gives $H(S|R_i)=1$, and $I(S:R_i)=1$. Hence, each neuron provides one bit of information about the stimulus when considered individually. Already here, we see something curious -- although each neuron has 1 bit about the stimulus, together they have only 2, not 3 bits. We can see the reason for this `vanishing bit' when considering responses from pairs of neurons, especially the pair $\{R_1,R_2\}$. 

What is the information in joint variables formed from pairs of neurons? If we first look at neurons 1 and 2 their responses to each stimulus are identical. Each of the neurons provides one bit of information about the stimulus. Even if we look at the two of them jointly  ($\{R_1,R_2\}$) we still get only one bit: $I(S: R_1,R_2)=1$. This is because the information carried by their responses is redundant. To see this, consider that we cannot decide between stimuli $s_1$ and $s_3$ if we get the result $(r_1=0,r_2=0)$, and we can also not decide between stimuli  $s_2$ and $s_4$ when observing $(r_1=1,r_2=1)$; other combinations of responses do not occur here. We see that neurons 1 and 2 have exactly the same information about the stimulus, and a measure of redundant information should yield the full 1 bit in this case (we will later see this intuitive argument again as the `Self Redundancy' axiom \citep{PaulWilliamsPID}).

To understand the concept of synergy, we next consider the output of responses $\{R_1, R_3\}$ from example neurons 1,3. We will transform these responses by a network that implements the mathematical \texttt{XOR} function, such that a downstream neuron $N$ at the output end of this \texttt{XOR}-network gets  activated only if there is one small bar on the screen (i.e. one of our neurons $R_1$ or $R_3$ gets activated, but not both), but neither for no stimulus nor for the long  bar (Figure \ref{fig:PIDneurons} C). We will now investigate the mutual information between $\{R_1, R_3\}$, $R_1$, $R_3$ and $N$. In this case the individual mutual informations of each neuron $R_1$, $R_3$ with the downstream neuron $N$ are zero ($I(N:R_i)=0$). However, the mutual information between these two neurons considered jointly and the downstream neuron $N$ is still 1 bit, because the response of $N$ is fully determined by its two inputs: $I(N:R_1,R_3)=1$.  Thus, there is only synergistic information between $R_1$ and $R_3$ about $N$. 

\begin{figure}
\begin{center}
\includegraphics[height=0.95\textheight]{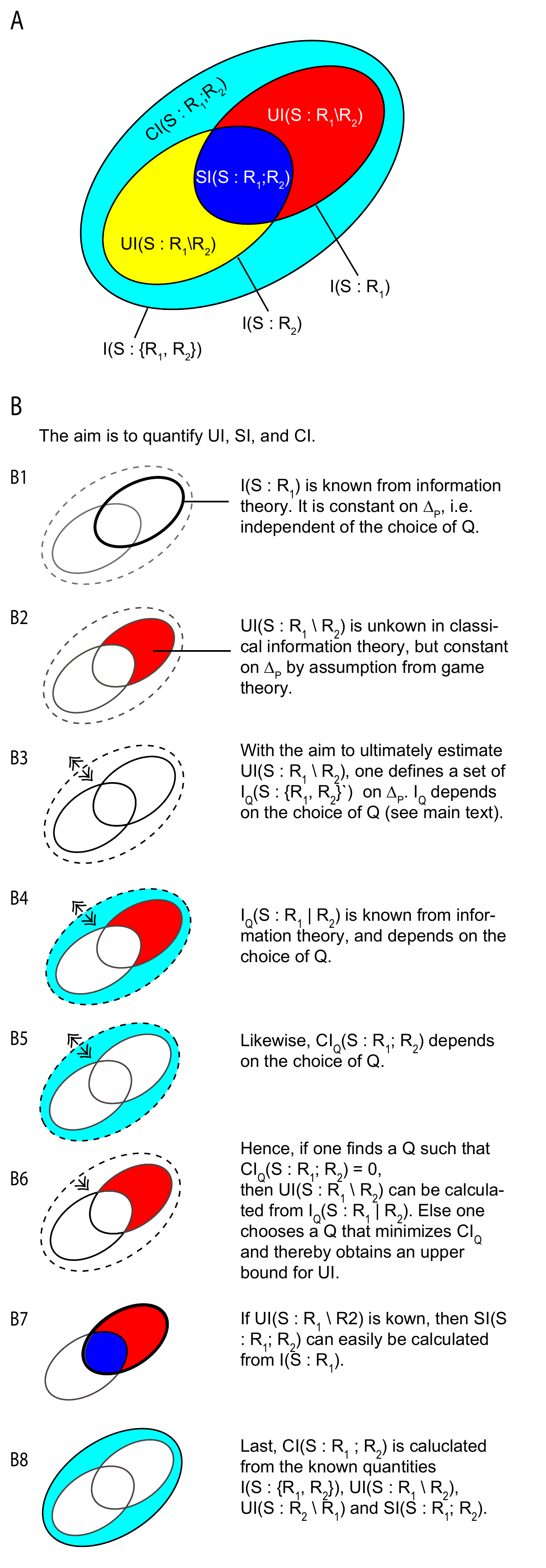}
\end{center}
\textbf{\refstepcounter{figure}\label{fig:PID} Figure \arabic{figure}.}{ Graphical depiction of the principle behind the definition of unique information in \citet{Bertschinger2014}. This figure is meant as a guide to the structure of the original work that should be consulted for the rigourous treatment of the topic.}
\end{figure}

We now introduce the mathematical framework of partial information decomposition that formalizes the intuition in the above examples, and consider a decomposition of the mutual information between a set of two right hand side, or input, variables $R_1$, $R_2$ and a left hand side variable, or output variable $S$, i.e. $I(S:R_1,R_2)$.
In general, for a decomposition of this mutual information into unique, redundant and synergistic information to make sense, the total information from any one variable, say $I(S:R_1)$ from $R_1$, should be decomposable into the unique information term $UI(S:R_1 \setminus R_2)$ and the redundant, or shared, information $SI(S:R_1;R_2)$ that both variables have about $S$:
\begin{equation}
  \label{eq:PID-2}
  \begin{split}
    I(S:R_1) &= SI(S:R_1;R_2) + UI(S:R_1\setminus R_2), \\
    I(S:R_2) &= SI(S:R_2;R_1) + UI(S:R_2\setminus R_1).
  \end{split}
\end{equation}
Similarly, the total information $I(S:R_1,R_2)$ from both variables should be decomposable into the two unique information terms $UI(S:R_1 \setminus R_2)$ and $UI(S:R_2 \setminus R_1)$ of each $R_i$ about $S$, the redundant, or shared, information $SI(S:R_1;R_2)$ that both variables have about $S$, and the synergistic, or complementary, information $CI(S:R_1;R_2)$ that can only be obtained by considering $\{R_1,R_2\}$ jointly:
\begin{equation}\label{eq:PID}
 I(S:R_1,R_2) = UI(S:R_1 \setminus R_2) + UI(S:R_2 \setminus R_1)+SI(S:R_1;R_2)+CI(S:R_1;R_2)~.
\end{equation}
\noindent Figure \ref{fig:PID}(A) shows this so called \emph{partial information decomposition} \citep{PaulWilliamsPID}. One sees that the redundant, unique and synergistic information cannot be obtained by simply subtracting classical mutual information terms. However, if we are given either a measure of redundancy, synergy or unique information, the other parts of the decomposition can be computed. Hence, classic information theory is insufficient for a partial information decomposition \citep{PaulWilliamsPID}, and a definition of either unique, redundant of synergistic information based on a choice of axioms is needed. A minimal requirement for such axioms, and measures satisfying them, is that that they should comply to our intuitive notion of what unique, redundant and synergistic information should be in some clear cut extreme cases, such as the examples above. The original set of axioms proposed for such a functional definition of redundant (and thereby also unique and synergistic) information comprises three axioms that all authors seem to agree on \citep{PaulWilliamsPID}:

\begin{enumerate}
 \item (Weak) Symmetry: The redundant information that variables $R_1$, $R_2$, $\ldots$, $R_n$ have about $S$  is symmetric under permutations of the variables $R_1$, $R_2$, $\ldots$, $R_n$.
 \item Self-redundancy: The redundant information that $R_1$ shares with itself about $S$ is just the mutual information $I(R_1:S)$
 \item Monotonicity: The redundant information that variables $R_1$, $R_2$, $\ldots$, $R_n$ have about $S$  is smaller or equal than the redundant information that variables $R_1$, $R_2$, $\ldots$, $R_{n-1}$ have about $S$. Equality holds if $R_{n-1}$ is a function of $R_n$.
\end{enumerate}

\noindent These three axioms also lead to global positivity, i.e. $SI(\cdot:\cdot) \ge 0$, $CI(\cdot:\cdot) \ge 0$ and $UI(\cdot:\cdot) \ge 0$ \citep{PaulWilliamsPID}. As said above, these axioms are uncontroversial, although some authors restrict them to only two input variables $R_1,~R_2$ as detailed below \citep{Rauh2014,Harder2013}. These axioms, however, are not sufficient to uniquely define a measure of either redundant, unique or synergistic information. Therefore, various additional axioms, or assumptions, have been proposed \citep{PaulWilliamsPID,Harder2013,Bertschinger2014,Lizier2013InfoMod,Griffith2014a,Griffith2014b,Timme2014} that are not all compatible with each other \citep{Bertschinger2012}.
Here we exemplarily discuss the recent choice of an assumption by \citet{Bertschinger2014} to define a measure of unique information, which is in fact equivalent to another formulation proposed by \citet{Griffith2014a}. The reasons for selecting this particular assumption are that at the time of writing it comes with the richest set of derived theorems, and that it has an appealing link to game theory and utility functions, and thus to measures of success of an agent or a BICS. We note at the outset that this is one of the measures that are defined only for two ``input'' variables $R_1$, $R_2$ and one ``output'' $S$ (although the $R_i$ themselves may be multivariate RVs). For more details on this restriction see \citet{Rauh2014}.

The basic idea of the definition by Bertschinger and colleagues comes from game theory and states that someone (say Alice) who has access to one input variable $R_1$ with unique information about an output variable $S$ must be able to prove that her variable has information not available in the other. To prove this, Alice can \emph{design} a bet on the output variable (by choosing a suitable utility function) so that someone else (say Bob) who has only access to the other input variable $R_2$ will on average loose this bet. Via some intermediate steps, this leads to the (defining) assumption that the unique information only depends on the two marginal probability distributions $P(s,r_1)$ and $P(s,r_2)$, but not on the exact full distribution $P(S,r_1,r_2)$. In other words, the unique information $UI$ should not change on the space $\Delta_P$ of probability distributions $Q$ that share these marginals with $P$:
\begin{multline}
  \Delta_{P} = \Big\{ Q\in\Delta : Q(S=s,R_1=r_1)=P(S=s,R_1=r_1)\\
  \text{ and }Q(S=s,R_2=r_2)=P(S=s,R_2=r_2)\text{ for all }s\in\mathcal{A}_S,r_1\in\mathcal{A}_{R_1},r_2\in\mathcal{A}_{R_2} \Big\}
\end{multline}
\noindent where $\Delta$ is the space of all probability distributions on the support of $S$, $R_1$, $R_2$.
This motivated to the following definition for a measure $\TUI$ of unique information:
\begin{equation}
  \TUI(S:R_1\setminus R_2) = \min_{Q\in\Delta_{P}} I_{Q}(S:R_1|R_2),
\end{equation}
\noindent where $I_{Q}(S:R_1|R_2)$ is a conditional mutual information computed with respect to the joint distribution~$Q(s,r_1,r_2)$ instead of $P(s,r_1,r_2)$. Note that this conditional mutual information $I_{Q}(S:R_1|R_2)$ does change on $\Delta_P$, and that only its minimum is a measure of the (constant) unique information (see Figure \ref{fig:PID}). As stated above, knowing one of the three parts $UI$, $SI$, $CI$ is enough to compute the others. Therefore, the matching definitions of measures for redundant ($\TSI$) and shared information ($\TCI$) are:%
\begin{align}
  \TSI(S:R_1;R_2) &= \max_{Q\in\Delta_{P}} CoI_{Q}(S:R_1;R_2), \\
  \TCI(S:R_1;R_2) &= I(S:R_1,R_2) - \min_{Q\in\Delta_{P}}I_{Q}(S:R_1,R_2).
\end{align}
\noindent where $CoI_{Q}(S;R_1;R_2)= I(S:R_1)-I_Q(S:R_1|R_2)$ is the so-called coinformation (equivalent to the redundancy minus the synergy) for the distribution~$Q(s,r_1,r_2)$.

Among the notable properties of the measures defined this way is the fact that they can be found by convex optimization, and that all three measures above have been explicitly shown to be positive. Moreover, \emph{the above measures are bounds for any definitions} of synergy $CI$, redundancy $SI$ and unique information $UI$ that satisfy equations \ref{eq:PID-2} and \ref{eq:PID}.
That is, it can be shown that:
 \begin{align*}
    UI(S:R_1\setminus R_2)  & \le \TUI(S:R_1\setminus R_2),  \\
    UI(S:R_2\setminus R_1)  & \le \TUI(S:R_2\setminus R_1),  \\
    SI(S:R_1;R_2)           & \ge \TSI(S:R_1;R_2),           \\
    CI(S:R_1;R_2)           & \ge \TCI(S:R_1;R_2),
  \end{align*}
\noindent holds \citep{Bertschinger2014}.

The field of information decomposition has seen a rapid development since the initial study of Williams and Beer, however, some major questions remain unresolved so far. Most importantly, the definitions above have acceptable properties, but apply only for the case of decomposing mutual information into contributions of two (sets of) input variables. The structure of such a decomposition for more than two inputs is an active area of research at the moment.

\section{Analyzing Distributed Computation in Neural Systems}
\label{sec:DistComp}

\subsubsection*{Analyzing neural coding and goal functions in a domain-independent way.}
The analysis of neural coding strategies presented above relies on our \emph{a priori}
knowledge of the set of task level (e.g. stimulus) features that is
encoded in neural responses at the implementation level.
If we have this knowledge, information theory will help us to link the
two levels. This is somewhat similar to the situation in cryptography
where we consider a code `cracked' if we obtain a human-readable plain
text message, i.e. we move from the implementation level (encrypted
message) to the task level (meaning). However, what happens if the plain text were in a language that one never heard of\footnote{See for example the Navajo code during World War Two that was never deciphered \citep{Fox_Chester_2014}.}? In this case, we would potentially crack the code without ever realizing it, as the plain text still has no meaning for us.

The sitution in neuroscience bears resemblance to this example in at last two respects: First, most neurons do not have direct access to any properties of the outside world, rather they receive nothing but input spike trains. All they ever learn and process must come from the structure of these input spike trains. Second, if we as researchers probe the system beyond early sensory or motor areas, we have little
knowledge of what is actually encoded by the neurons deeper inside the system. As a result proper stimulus sets get hard to choose. In this case, the gap between the task- and the implementation level may actually become too wide for meaningful analyses, as noticed
recently by \citet{Carandini2012}.

Instead of relying on descriptions of the outside world (and thereby involve the task level), we may take the point of view that information processing in a neuron is nothing but the transformation of input spike trains to output spike trains. We may then try  to use information theory to link the implementation and algorithmic level, by retrieving a `footprint' of the information processing carried out by a neural circuit. This approach only builds on a very general agreement that neural systems perform at least \emph{some kind of} information processing. This information processing can be decomposed into the component processes of (1) information storage, (2) information transfer, and (3) information modification. A decomposition of this kind had
already been formulated by Turing (see \citet{Langton1990}), and was recently formalized by \citet{Lizier2014c} (see also \citet{LizierBook}):

\begin{itemize}
 \item \textbf{Information storage} quantifies the information contained
in the past state variable $\mathbf{Y_{t-1}}$ of a process that is used by the process at the next RV at $t$, $Y_{t}$ \citep{Lizier2012a}. This relatively abstract definition means
that we will see at least a part of the past information again in the
future of the process, but potentially transformed. Hence, information
storage can be naturally quantified by a mutual information between the
past and the future\footnote{We consider ourselves having information up to time $t-1$, predicting the future values at $t$.} of a process.

 \item \textbf{Information transfer} quantifies the information
contained in the past state variables $\mathbf{X_{t-u}}$ of one source
process \texttt{X} that can be used to predict information in the future variable $Y_t$ of a target process \texttt{Y}, in the context of the past state variables $\mathbf{Y_{t-1}}$ of the target process \citep{Schreiber2000,pal01.1,VWL+11}.

  \item\textbf{Information modification } quantifies the combination of
information from various source processes into a new form that is not
(trivially) predictable from any subset of these source processes
\citep{Lizier2013InfoMod,Lizier2010a}.
\end{itemize}

Based on Turing's general decomposition of information processing \citep{Langton1990}, Lizier and colleagues recently proposed an information theoretic framework to quantify distributed computations in
terms of all three component processes \emph{locally}, i.e. for each part of the system (e.g. neurons or brain areas) and each time step \citep{Lizier2008,Lizier2010a,Lizier2012a}. This framework is called
\emph{local information dynamics} and has been successfully applied to
unravel computation in swarms \citep{Wang2011}, in Boolean networks
\citep{Lizier2011}, and in neural models \citep{Boedecker2012} and data \citep{Wibral2014VSDLAIS} (also see \secRef{AppEx} for details on these example applications).

In the following we present both global and local measures of
information transfer, storage and modification, beginning with the well
established measures of information transfer and ending with the highly
dynamic field of information modification.

\subsection{Information Transfer}
The analysis of information transfer was formalized initially by
\citet{Schreiber2000} and \citet{pal01.1}, and
has seen a rapid surge of interest in neuroscience\footnote{
\citep{VWL+11,WRR+11,pal01.1,Vakorin2010a,Vakorin2009,Vakorin2011a,Chavez2003,AM10,Barnett2009,BSL+10,GNM+09,Gourevitch2007,Lizier2011a,LLP10,NJF+10,SGT+09,SL09,Leistritz2006c,Hadjipapas2005,NJF+10,Lindner2011,Battaglia2012a,Stetter2012,Ito2011,BSL+10,LiAndLi,Besserve2010,Buehlmann2010,zubler2014,yamaguti2014,varon2014interictal,untergehrer2014,thivierge2014,shimono2014,rowan2014electrostimulation,razak2014,porta2014effect,orlandi2014,montalto2014,van2014functional,mcauliffe2014,Marinazzo2014,marinazzo2014information,marinazzo2014ScalpEEG,liu2014new,kawasaki2014,chicharro2014parametric,butail2014information,bedo2014fast,battaglia2014function}  }
and general physiology\footnote{\citep{Faes2006a,Faes2011c,Faes2011b,faes2014conditional,faes2014information,faes2014LagEnt}}.

\subsubsection{Definition}
Information transfer from a process \texttt{X} (the \emph{source}) to
another process \texttt{Y} (the \emph{target}) is measured by the
transfer entropy (TE) functional\footnote{A functional maps from the
relevant probability distribution (i.e. functions) to the real numbers.
In contrast, an estimator maps from empirical data, i.e. a set of real
numbers, to the real numbers.} \citep{Schreiber2000}:
\begin{align}
\label{eq:TEfunc}
  TE(X_{t-u} \rightarrow Y_t)&=I(\mathbf{X}_{t-u}: Y_t | \mathbf{Y}_{t-1})\\
  &=\sum_{y_t \in \mathcal{A}_{Y_t}, \mathbf{y}_{t-1}\in
\mathcal{A}_{\mathbf{Y}_{t-1}},\mathbf{x}_{t-u}\in \mathcal{A}_{\mathbf{X}_{t-u}}}
  p(y_t, \mathbf{y}_{t-1},\mathbf{x}_{t-u})\log \frac {p(y_t |
\mathbf{y}_{t-1},\mathbf{x}_{t-u})}{p(y_t | \mathbf{y}_{t-1})}
  ~,
\end{align}
\noindent where $I( \cdot:\cdot| \cdot )$ is the conditional mutual
information, $Y_t$ is the RV of process \texttt{Y} at time $t$,
and $\mathbf{X}_{t-u},\mathbf{Y}_{t-1}$ are the past state-RVs of
processes \texttt{X} and \texttt{Y}, respectively. The delay variable $u$ in $\mathbf{X}_{t-u}$ indicates that the past state of the source is to be taken $u$ time steps into the past to account for a potential physical interaction delay between the processes. This parameter need not be chosen \emph{ad hoc}, as it was recently proven for bivariate systems that the above estimator
is maximized if the parameter $u$ is equal to the true delay $\delta$ of the information transfer from \texttt{X} to \texttt{Y} \citep{Wibral2013}. This relationship allows one to estimate the \emph{true} interaction delay $\delta$ from data by simply scanning the assumed delay $u$:
\begin{equation}
\label{eq:delay}
\delta = \underset{u}{\text{argmax}}~[TE\left(\textbf{X}_{t-u}\rightarrow
Y_t,u\right)]
\end{equation}

The TE functional can be linked to Wiener-Granger type causality \citep{Wiener1956,Granger1969,Barnett2009}.  More precisely, for systems with \emph{jointly} Gaussian variables, transfer entropy is equivalent\footnote{To a constant factor of 2.} to \emph{linear} Granger causality (see \citet{Barnett2009} and references therein). However, whether the assumption of jointly Gaussian variables is appropriate in a neural setting must be checked carefully for each case (note that Gaussianity of each marginal distribution is not sufficient). In fact, EEG source signals were found to be non-Gaussian \citep{Wibral2008}

We also note that TE has recently been given a thermodynamic interpretation by \citet{prokopenko2014TransientLimits}.

\subsubsection{Transfer Entropy Estimation}
When the probability distributions entering eq. \ref{eq:TEfunc} are known (e.g. in an analytically tractable neural model), TE can be computed directly. However, in most cases the probability distributions have to be derived from data. When probabilities are estimated naively from the data via couting, and when these estimates are then used to compute information theoretic quantities such as the transfer entropy, we speak of a ``plug in'' estimator. Indeed such plug in estiamtors have been used in the past, but they come with serious bias problems \citep{PSM+07}. Therefore, newer approaches to TE estimation rely on a more direct estimation of the entropies that TE can be decomposed into
\citep{Kraskov2004,Gomez-Herrero2010,VWL+11,Wibral2014a}. These estimators still suffer from bias problems but to a lesser degree \citep{Kraskov2004}. We therefore restrict our presentation to these approaches.

Before we can proceed to estimate TE we will have to reconstruct the states of the processes (see \secRef{states}). One approach to state reconstruction is time delay embedding \citep{Takens1981}. It uses past variables $X_{t-n\tau},~n=1,2,\ldots$ that are spaced in time by an interval $\tau$. The number of these variables and their optimal spacing can be determined using established criteria
\citep{Lindner2011,Faes2012,Small2004,Ragwitz2002}. The realizations of the states variables can be represented as vectors of the form:
\begin{equation}
\label{eq:del_vec}
\begin{split}
\mathbf{x}^{d}_{t}=(x_{t},x_{t-\tau},x_{t-2\tau},...\\
 \quad ,x_{t-\left(d-1\right)\tau}) ~,
\end{split}
\end{equation}
\noindent where $d$ is the dimension of the state vector. Using this vector notation, the transfer entropy estimator writes:

\begin{equation}
\label{eq:TE_p}
\begin{split}
TE_{SPO}\left(\textbf{X}_{t-u}\rightarrow Y_t,u\right)=
\sum_{y_{t},\mathbf{y}^{d_{y}}_{t-1},\mathbf{x}^{d_{x}}_{t-u}} p\left(
y_{t}, \mathbf{y}^{d_{y}}_{t-1}, \mathbf{x}^{d_{x}}_{t-u}  \right) \\
\log \frac{p\left( y_{t} | \mathbf{y}^{d_{y}}_{t-1},
\mathbf{x}^{d_{x}}_{t-u} \right)}{p\left(y_{t} |
\mathbf{y}^{d_{y}}_{t-1}\right)} ~,
\end{split}
\end{equation}

\noindent where the subscript $SPO$ (for \underline{s}elf
\underline{p}rediction \underline{o}ptimal) is a reminder that the past
states of the target, $\mathbf{y}^{d_{y}}_{t-1}$, have to be constructed such that conditioning on them is optimal in the sense of taking the active information storage in the target correctly into account \citep{Wibral2013}: If one were to condition on
$\mathbf{y}^{d_{y}}_{t-w}$ with $w \neq 1$, instead of
$\mathbf{y}^{d_{y}}_{t-1}$, then the self prediction for $Y_t$ would not
be optimal and the transfer entropy would be overestimated.

We can rewrite equation \ref{eq:TE_p} using a representation in the form
of four entropies\footnote{For continuous-valued RVs, these
entropies are differential entropies.} $H(\cdot)$, as:
\begin{equation}
\label{eq:TE_4H}
\begin{split}
TE_{SPO}\left(\textbf{X}_{t-u}\rightarrow
Y_t,u\right) = H\left(\mathbf{y}^{d_{y}}_{t-1},
\mathbf{x}^{d_{x}}_{t-u}  \right) - H\left(y_{t},
\mathbf{y}^{d_{y}}_{t-1}, \mathbf{x}^{d_{x}}_{t-u}  \right)\\
 + H\left( y_{t}, \mathbf{y}^{d_{y}}_{t-1} \right) - H\left(
\mathbf{y}^{d_{y}}_{t-1} \right) \, .
\end{split}
\end{equation}

\noindent Entropies can be estimated
efficiently by nearest-neighbor techniques. These techniques exploit the
fact that the distances between neighboring data points in a given
embedding space are inversely related to the local probability density: the higher
the local probability density around an observed data point 
the closer are the next neighbors. Since next neighbor estimators are data efficient \citep{koz87.1,victor02.1} they allow to estimate entropies in high-dimensional spaces from limited real data.

Unfortunately, it is problematic to estimate TE by simply applying a
naive nearest-neighbor estimator for the entropy, such as the
Kozachenko-Leonenko estimator \citep{koz87.1}, separately to each of the
terms appearing in equation \ref{eq:TE_4H}. The reason is that the
dimensionality of the state spaces involved in equation \ref{eq:TE_4H}
differ largely across terms -- creating bias problems. These are overcome by the Kraskov-St\"ogbauer-Grassberger (KSG) estimator that fixes the number of neighbors $k$ in the
highest dimensional space (spanned here by $y_{t},~\mathbf{y}^{d_{y}}_{t-1},~\mathbf{x}^{d_{x}}_{t-u}$) and by projecting the resulting distances to the lower dimensional spaces as the range to look for additional neighbors there
\citep{Kraskov2004}. After adapting this technique to the TE formula
\citep{Gomez-Herrero2010}, the suggested estimator can be written as:
\begin{align}
\label{eq:TE_kras}
TE_{SPO}\left(X\rightarrow Y,u\right) = \psi\left(k\right)+\Big\langle
\psi\left(n_{\mathbf{y_{t-1}^{d_{y}}}}+1\right) 
 -\psi\left(n_{y_{t}\mathbf{y_{t-1}^{d_{y}}}}+1\right)
-\psi\left(n_{\mathbf{y_{t-1}^{d_{y}}x_{t-u}^{d_{x}}}}+1\right)
\Big\rangle_{t} 
\, ,
\end{align}

\noindent where $\psi$ denotes the digamma function, the angle
brackets ($\langle \cdot \rangle_{t}$) indicate averaging over
time for stationary systems, or over an ensemble of replications for non-stationary ones, and $k$ is the number of nearest neighbors used for the estimation.
$n_{(\cdot)}$ refers to the number of neighbors which are within a hypercube that defines the search range around a state vector. As described above, the size of the hypercube in each of the marginal spaces is defined based on the distance to the $k$-th nearest neighbor in the highest dimensional space.

\subsubsection{Interpretation of transfer entropy as a measure at the
algorithmic level.}
TE describes computation at the algorithmic level, not at the level of a physical dynamical system. As such it is not optimal for inference about \emph{causal} interactions - although it has been used for this purpose in the past. The fundamental reason for this is that information transfer relies on causal interactions, but causal interactions do not necessarily lead to nonzero information transfer \citep{Ay2008,Lizier2010,Chicharro2012}. Instead, causal interactions may serve active information storage alone (see next section), or force two systems into identical synchronization, where information transfer becomes effectively zero. This might be summarized by stating that transfer entropy is limited to effects of a causal interaction from a source to a target process that are unpredictable given the past of the target process alone. In this sense, TE may be seen as quantifying causal interactions currently \emph{in use for the communication aspect} of distributed computation. Therefore, one may say that TE measures \emph{predictive}, or \emph{algorithmic information transfer}.

A simple thought experiment may serve to illustrate this point: When one plays an unknown record, a chain of causal interactions serve the transfer of information about the music from the record to your brain. Causal interactions happen between the record's grooves and the needle, the magnetic transducer system behind the needle, and so on, up to the conversion of pressure modulations to neural signals in the cochlea that finally activate your cortex. In this situation, there undeniably is information transfer, as the information read out from the source, the record, at any given moment is not yet known in the target process, i.e. the neural activity in the cochlea. However, this information transfer ceases if the record has a crack, making the needle skip and repeat a certain part of the music. Obviously, no new information is transferred which under certain mild conditions is equivalent to no information transfer at all. Interestingly, an analysis of TE between sound and cochlear activity will yield the same result: The repetitive sound leads to repetitive neural activity (at least after a while). This neural activity is thus predictable by it's own past, under the condition of vanishing neural `noise', leaving no room for a prediction improvement by the sound source signal. Hence, we obtain a TE of zero, which is the correct result from a conceptual point of view. Remarkably, at the same time the chain of causal interactions remains practically unchanged. Therefore, a causal model able to fit the data from the original situation will have no problem to fit the data of the situation with the cracked record, as well. Again, this is conceptually the correct result, but this time from a causal point of view.

The difference between an analysis of information transfer in a computational sense and causality analysis based on interventions has demonstrated convincingly in a recent study by \citet{Lizier2010}. The same authors also demonstrated why an analysis of information transfer can yield better insight than the analysis of causal interactions if the \emph{computation} in the system is to be understood. The difference between causality and information transfer is also reflected in the fact that a single causal structure can support diverse pattern of information transfer (\emph{functional multiplicity}), and the same pattern of information transfer can be realized with different causal structures (\emph{structural degeneracy}) as shown by \citet{Battaglia2014}.

\subsubsection{Local information transfer}
As transfer entropy is formally just a conditional mutual information, we can obtain the corresponding local conditional mutual information (equation \ref{eq:lcmi}) from equation \ref{eq:TE_p}. This quantity is called the local transfer entropy \citep{Lizier2008}. For realizations $x_t,y_t$ of two processes \texttt{X}, \texttt{Y} at time $t$ it reads:
\begin{equation}
\label{eq:lte}
te\left(\mathbf{X}_{t-u}=\mathbf{x}_{t-u}\rightarrow Y_t=y_t\right)= \log \frac{p\left( y_{t} | \mathbf{y}^{d_{y}}_{t-1},
\mathbf{x}^{d_{x}}_{t-u} \right)}{p\left(y_{t} |
\mathbf{y}^{d_{y}}_{t-1}\right)} ~,
\end{equation}

\noindent As said earlier in the section on basic information theory, the use of local information measures does not eliminate the need for an appropriate estimation of the probability distributions involved. Hence, for a non-stationary process these distributions will still have to be estimated via an ensemble approach for each time point for the RVs involved -- e.g. via physical replications of the system, or via enforcing cyclostationarity by design of the experiment. 

The analysis of local transfer entropy has been applied with great success in the study of cellular automata to confirm the conjecture that certain coherent spatio-temporal structures traveling through the network are indeed the main carriers of information transfer \citep{Lizier2008} (see further discussion at \secRef{CAs}).
Similarly, local transfer entropy has identified coherent propagating wave structures in flocks as information cascades \citep{Wang2012} (see \secRef{Swarms}), and indicated impending synchronization amongst coupled oscillators \citep{Ceguerra2011}.

\subsubsection{Common Problems and solutions}
Typical problems in TE estimation encompass (1) finite sample bias, (2) the presence of non-stationarities in the data, and (3) the need for multivariate analyses. In recent years all of these problems have been addressed at least in isolation, as summarized below:

\begin{enumerate}
 \item Finite sample bias can be overcome by statistical testing using surrogate data, where the observed realizations $ y_{t}, \mathbf{y}^{d_{y}}_{t-1}, \mathbf{x}^{d_{x}}_{t-u} $ of the RVs $ Y_{t}, \mathbf{Y}^{d_{y}}_{t-1}, \mathbf{X}^{d_{x}}_{t-u} $ are reassigned to other RVs of the process, such that the temporal order underlying the information transfer is destroyed (for an example see the procedures suggested in \citep{Lindner2011}). This reassignment should conserve as many data features of the single process realizations as possible.

\item As already explained in the section on basic information theory above, non-stationary random processes in principle require that the necessary estimates of the probabilities in equation \ref{eq:TEfunc} are based on physical replications of the systems in question. Where this is impossible, the experimenter should design the experiment in such a way that the processes are repeated in time. If such cyclostationary data are available, then TE should be estimated using ensemble methods as described in \citep{Gomez-Herrero2010} and implemented in the TRENTOOL toolbox \citep{Lindner2011,Wollstadt2014}.

\item So far, we have restricted our presentation of transfer entropy estimation to the case of just two interacting random processes \texttt{X} , \texttt{Y}, i.e. a bivariate analysis. In a setting that is more realistic for neuroscience, one deals with large networks of interacting processes \texttt{X}, \texttt{Y}, \texttt{Z}, \ldots~. In this case various complications arise if the analysis is performed in a bivariate manner. For example a process \texttt{Z} could transfer information with two different delays $\delta_{Z\rightarrow X}$, $\delta_{Z\rightarrow Y}$ to two other processes  \texttt{X}, \texttt{Y}. In this case, a pairwise analysis of transfer entropy between $X$, $Y$ will yield an apparent information transfer from the process that receives information from \texttt{Z} with the shorter delay to the one that receives it with the longer delay (common driver effect). A similar problem arises if information is transferred first from a process \texttt{X} to \texttt{Y}, and then from \texttt{Y} to \texttt{Z}. In this case, a bivariate analysis will also indicate information transfer from \texttt{X} to \texttt{Z} (cascade effect). Moreover, two sources may transfer information purely synergistically, i.e. the transfer entropy from each source alone to the target is zero, and only considering them jointly reveals the information transfer\footnote{ Again, cryptography may serve as an example here. If an encrypted message is received, there  will be no discernible information transfer from encrypted message to plain text without the key. In the same way, there is no information transfer from the key alone to the plain text. It is only when encrypted message and key are combined that the relation between the combination of encrypted message and key on the one side and the plain text on the other side is revealed.}.

From a mathematical perspective this problem seems to be easily solved by introducing the \emph{complete transfer entropy} \citep{Lizier2008,Lizier2010a}, which is defined in terms of a \emph{conditional transfer entropy} \citep{Lizier2008,Lizier2010a}:
\begin{equation}
 \label{eq:CondTE}
 TE\left(\mathbf{X_{t-u}}\rightarrow Y_t | \mathbf{Z^-} \right)= \sum_{y_{t},\mathbf{y_{t-1}},\mathbf{x_{t-u}},\mathbf{z^-}} p\left( y_{t}, \mathbf{y_{t-1}}, \mathbf{x_{t-u}},  \mathbf{z^-} \right) \\ \log \frac{p\left( y_{t} |\mathbf{y_{t-1}},\mathbf{x_{t-u}},\mathbf{z^-} \right)}{p\left( y_{t}|\mathbf{y_{t-1}}, \mathbf{z^-}  \right)} ,
\end{equation}
where the state-RV $\mathbf{Z^-}$ is a collection of the past states of \emph{one or more} processes in the network other than \texttt{X}, \texttt{Y}.
We label \eq{CondTE} a complete transfer entropy $TE^{(c)}\left(\mathbf{X_{t-u}}\rightarrow Y_t \right)$ when we take $\mathbf{Z^-} = \mathbf{V^-}$, the set of \emph{all} processes in the network other than \texttt{X}, \texttt{Y}.

However, even for small networks of random processes the joint state space of the variables $Y_t,\mathbf{Y_{t-1}},\mathbf{X_{t-u}},\mathbf{V^-}$ may become intractably large from an estimation perspective. Moreover, the problem of finding all information transfers in the network, either from single sources variables into the target or synergistic transfer from collections of source variables to the target, is a combinatorial problem, and can therefore typically not be solved in a reasonable time.

Therefore, \citet{Faes2012}, \citet{LizierMPI}, and \citet{Stramaglia2012} suggested to analyze the information transfer in a network iteratively, selecting information sources for a target in each iteration either based on magnitude of apparent information transfer \citep{Faes2012}, or its significance \citep{LizierMPI,Stramaglia2012}. In the next iteration, already selected information sources are added to the conditioning set ($\mathbf{Z^-}$ in equation \ref{eq:CondTE}), and the next search for information sources is started. The approach of Stramaglia and colleagues is particular here in that the conditional mutual information terms are computed at each level as a series expansion, following a suggestion by \citet{Bettencourt2008}. This allows for an efficient computation as the series may truncate early, and the search can proceed to the next level. Importantly, these approaches also consider synergistic information transfer from more than one source variable to the target. For example, a variable transferring information purely synergistically with $\mathbf{Z^-}$  maybe included in the next iteration, given that that the other variables it transfers information with are already in the conditioning set $\mathbf{Z^-}$. However, there is currently no explicit  indication in the approaches of \citet{Faes2012},  \citet{LizierMPI} as to whether multivariate information transfer from a set of sources to the target is in fact synergistic; in addition, redundant links will not be included. In contrast, both redundant and synergistic multiplets of variables transferring information into a target may be identified in the approach of \citet{Stramaglia2012} by looking at the sign of the contribution of the multiplet. Unfortunately there is also the possibility of cancellation if both types of multivariate information (redundant, synergistic) are present.
\end{enumerate}

\subsection{Active Information Storage}
Before we present explicit measures of active information storage, a few comments may serve to avoid misunderstanding. Since we analyze neural \emph{activity} here, measures of active information storage are concerned with information stored in this activity -- rather than in synaptic properties, for example.\footnote{See the distinction made between passive storage in synaptic properties and active storage in dynamics by \citet{Zipser1993}.} As laid out above, storage is conceptualized here as a mutual information between past and future states of neural activity. From this it is clear that there will not be much information storage if the information contained in the future states of neural activity is low in general. If, on the other hand these future states are rich in information but bear no relation to past states, i.e. are unpredictable, again information storage will be low. Hence, large information storage occurs for activity that is rich in information but, at the same time, predictable.

Thus, information storage gives us a  way to define the predictability of a process  that is independent of the prediction error: information storage quantifies how much future information of a process can be predicted from its past, whereas the  prediction error measures how much information can not be predicted. If both are quantified via information measures, i.e. in bits, the error and the predicted information add up to the total amount of information in a random variable of the process. Importantly, these two measures may lead to quite different views about the predictability of a process. This is because the total information can vary considerably over the process, and the predictable and the unpredictable information may thus vary almost independently. This is important for the design of BICS that use predictive coding strategies.

Before turning to the explicit definition of measures of information storage it is worth considering which temporal extent of 'past' and 'future' states we are interested in: Most globally, \emph{predictive information} \citep{Bialek2001} or \emph{excess entropy} \citep{Crutchfield1982,Grassberger1986,Crutchfield2003a} is the mutual information between the \emph{semi-infinite} past and \emph{semi-infinite} future of a process before and after time point $t$. In contrast, if we are interested in the information currently used for the \emph{next step} of the process, the mutual information between the \emph{semi-infinite} past and the next step of the process, the \emph{active information storage} \citep{Lizier2012a} is of greater interest. Both measures are defined in the next paragraphs.

\subsubsection{Predictive information / Excess entropy}
Excess entropy is formally defined as:
\begin{equation}
\label{eq:ExcessEntDef}
E_{X_t}=\underset{k\rightarrow \infty}{\lim} I(\mathbf{X}^{k-}_t:\mathbf{X}^{k+}_t)
\end{equation}
\noindent where $\mathbf{X}^{k-}_t=\{\textbf{X}_t, \textbf{X}_{t-1},\ldots,\textbf{X}_{t-k+1}  \}$, and $\mathbf{X}^{k+}_t=\{\textbf{X}_{t+1},\ldots,\textbf{X}_{t+k}  \}$ indicate collections of the past and future $k$ variables of the process \texttt{X}.\footnote{In principle these could harness embedding delays, as defined in \eq{del_vec}.} These collections of RVs ($\mathbf{X}^{k-}_t$,$\mathbf{X}^{k+}_t$) in the limit $k\rightarrow \infty$ span the semi-infinite past and future, respectively. In general, the mutual information in equation \ref{eq:ExcessEntDef} has to be evaluated over multiple realizations of the process. For stationary process, however, $E_{X_t}$ is not time-dependent, and equation \ref{eq:ExcessEntDef} can be rewritten as an average over time points $t$ and computed from a single realization of the process -- at least in principle (we have to consider that the process must run for an infinite time to allow the limit $\underset{k\rightarrow \infty}{\lim}$ for all $t$):
\begin{equation}
\label{eq:ExcessEntDefStat}
E_X=\langle \underset{k\rightarrow \infty}{\lim} i(\mathbf{x}^{k-}_t:\mathbf{x}^{k+}_t)  \rangle_t
\end{equation}
\noindent where $i(\cdot:\cdot)$ is the local mutual information from equation \ref{eq:lmi}, and $\mathbf{x}^{k-}_t,~\mathbf{x}^{k+}_t$ are realizations of $\mathbf{X}^{k-}_t,~\mathbf{X}^{k+}_t$.  The limit of $ k\rightarrow \infty $ can be replaced by a finite $k_{\max}$ if a  $k_{\max}$ exists such that conditioning on $\mathbf{X}^{k_{\max}-}_t$ renders $\mathbf{X}^{k_{\max }+}_t$ conditionally independent of any $X_{l}$ with $l\leq t-k_{\max}$. 

Even if the process in question is non-stationary, we may look at values that are local in time as long as the probability distributions are derived appropriately (see 
\secRef{EstProb}): 

\begin{equation}
 \label{eq:ExcessEntDefLocal}
e_{X_t}=\lim_{k\rightarrow \infty}{i(\mathbf{x}^{k-}_t:\mathbf{x}^{k+}_t)}~.
\end{equation}

\subsubsection{Active Information Storage}
From a perspective of the dynamics of information processing, we might not be interested in information that is used by a process at some time far in the future, but at the next point in time, i.e. information that is said to be `currently in use' for the computation of the next step (the realization of the next RV) in the process \citep{Lizier2012a}. To quantify this information, a different mutual information is computed, namely the \emph{active information storage} (AIS):
\begin{equation}
 \label{AIS}
A_{X_t}= \underset{k\rightarrow \infty}{\lim} I(\mathbf{X}^{k-}_{t-1}:X_{t})
\end{equation}
\noindent Again, if the process in question is stationary then $A_{X_t}=\mathtt{const.}=A_X$ and the expected value can be obtained from an average over time -- instead of an ensemble of realizations of the process -- as:
\begin{equation}
 \label{AISStat}
 A_X=\langle \underset{k\rightarrow \infty}{\lim} i(\mathbf{x}^{k-}_{t-1}:x_{t})  \rangle_t
\end{equation}
\noindent which can be read as an average over local active information storage (LAIS) values $a_{X_t}$:
\begin{align}
 \label{AISStatloc}
 A_X&=\langle a_{X_t}\rangle_t\\
 a_{X_t}&=\underset{k\rightarrow \infty}{\lim} i(\mathbf{x}^{k-}_{t-1}:x_{t}) ~.
\end{align}

\noindent Even for nonstationary processes we may investigate local active storage values, given the corresponding probability distributions are properly obtained from an ensemble of realizations of $X_t$, $\mathbf{X}^{k-}_{t-1}$:
\begin{align}
 \label{AISNonStatloc}
 a_{X_t}&=\underset{k\rightarrow \infty}{\lim} i(\mathbf{x}^{k-}_{t-1}:x_{t}) ~.
\end{align}

\noindent Again, the limit of $ k\rightarrow \infty $ can be replaced by a finite $k_{\max}$ if a  $k_{\max}$ exists such that conditioning on $\mathbf{X}^{k_{\max}}_{t-1}$ renders $X_{t}$ conditionally independent of any $X_{l}$ with $l\leq t-k_{\max}$ (see equation \ref{eq:embedding1}).

\subsubsection{Interpretation of information storage as a measure at the algorithmic level}
As laid out above information storage is a measure of the amount of information in a process that is predictable from its past. As such it quantifies for example how well activity in one brain area $A$ can be predicted by another area, e.g. by learning its statistics. Hence, questions about information storage arise naturally when asking about the generation of predictions in the brain, e.g.\,in predictive coding.

\subsection{Combining the analysis of local active information storage and local transfer entropy}
The two measures of local active information storage and local transfer entropy introduced in the preceding section may be fruitfully combined by pairing storage and transfer values at each point in time and for each agent. The resulting space has been termed the ``local information dynamics state space'' and has been used to investigate the computational capabilities of cellular automata, by pairing $a(y_{j,t})$ and $te\left(x_{i,t-1}\rightarrow y_{j,t}\right)$ for each pair of sources and targets $x_i$, $y_j$ at each time point \citep{Lizier2012g}. 

Here, we suggest that this concept may be used to disentangle various neural processing strategies. Specifically we suggest to pair the sum\footnote{More complex ways of combining incoming active information storage are conceivable.} over all local active information storage in the \emph{inputs} $x_i$ of a target $y_j$ (at the relevant delays $u_i$, obtained from an analysis of transfer entropy \citep{Wibral2013}) with the sum of outgoing local information transfers from this target to further targets $z_k$, for each agent $y_j$ and each time point $t$:
\begin{align}
 \left( \sum_{x_i} a(x_{i,t-u_{i}})~,~ \sum_{ z_{k}} te(y_{j,t}\rightarrow z_{k,t+u_k}) \right)\\
 \intertext{where sources $x_i$ and second order targets $z_k$ are defined by the conditions:}\\
 te(x_{i,t-{u_i}}\rightarrow y_{j,t}) \neq 0,~\forall x_{i,t-{u_i}}\\
 te(y_{j,t}\rightarrow z_{k,t+u_k}) \neq 0,~\forall z_{k,t+{u_k}}.
\end{align}

\begin{figure}[h!]
\begin{center}
\includegraphics[width=7cm]{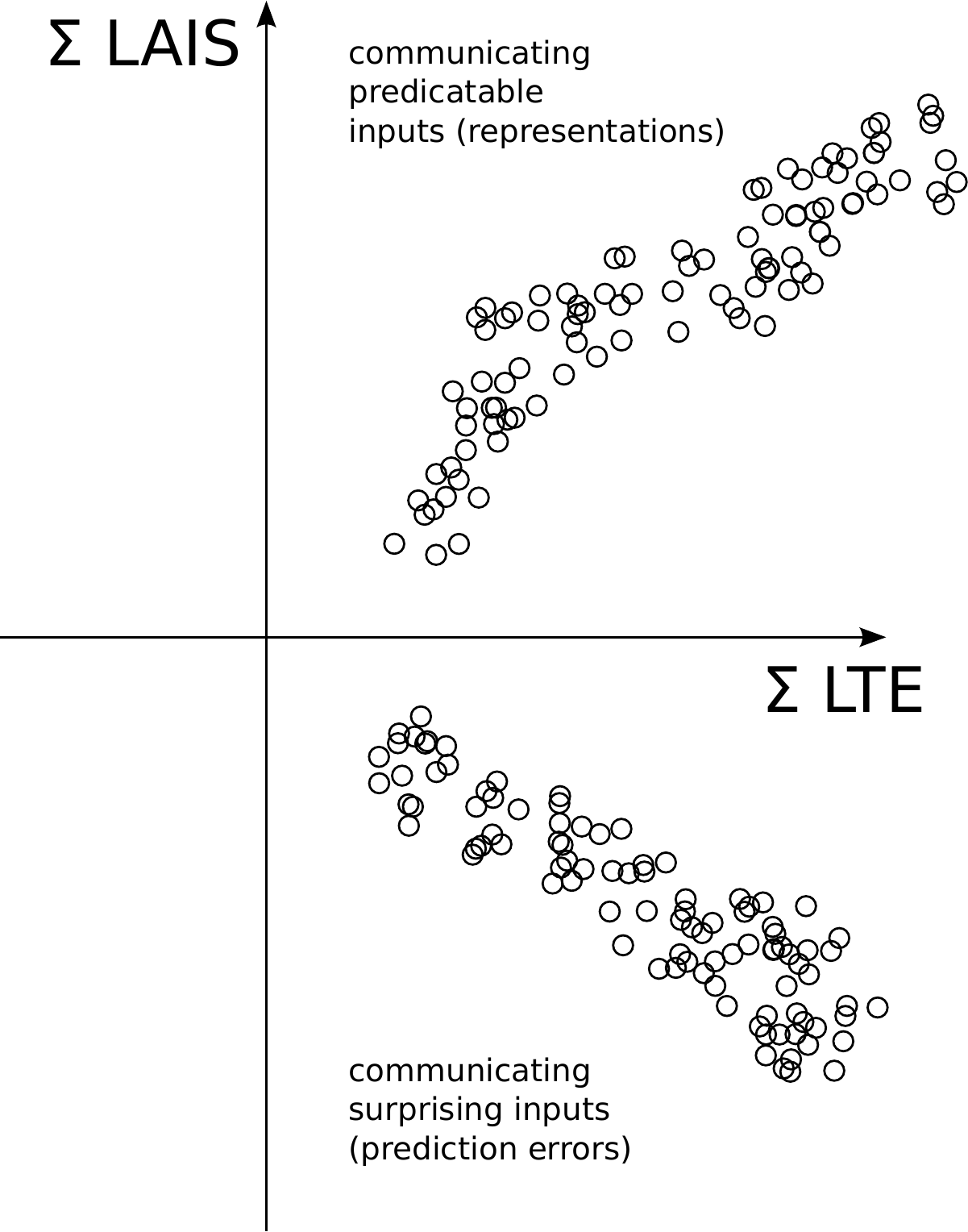}
\end{center}
\textbf{\refstepcounter{figure}\label{fig:infostatespace} Figure \arabic{figure}.}{ Various information processing regimes in the information state space. $\Sigma~LAIS$ = sum of local active information storage in input, $\Sigma~ LTE$ = sum of outgoing local transfer entropy. Each dot represents these values for one agent and time step.}
\end{figure}

\noindent The resulting point set set can be used to answer the important question, whether the aggregate outgoing information transfer of an agent is high either for predictable or for surprising input. The former information processing function amounts a sort of filtering, passing on reliable (predictable) information, and would be linked to something reliable being represented in activity. The latter information processing function is a form of prediction error encoding, where high outgoing information transfer is triggered when surprising, unpredictable information is received (also see Figure \ref{fig:infostatespace}).

Note that for this type of analysis recordings of at least \emph{triplets} of connected agents are necessary. This may pose a considerable challenge in experimental neuroscience, but may be extremely valuable to disentangle the information processing goal functions of the various cortical layers for example. This type of analysis will also be valuable to understand the information processing in evolved BICS, as in these systems the availability of data from triplets of agents is no problem.

\section{Information Modification and its Relation to Partial Information Decomposition}
\label{sec:InfoMod}
\citet{Langton1990} described information modification as an interaction between transmitted and/or stored information that results in a modification of one or the other. Attempts to define information modification more rigorously implemented this basic idea. First attempts at defining a quantitative measure of information modification resulted in a heuristic measure termed \emph{local separable information} \citep{Lizier2010a}, where  the local active information storage and the sum over all pairwise local transfer entropies into the target was taken:
\begin{equation}
\label{eq:sepInfo}
 s_{X_t}=a_{X_{t}}+\sum_{\mathbf{Z}_{t^-,i} \in \mathbf{V}_{X_{t}} \backslash \mathbf{X}_{t-1}} i(x_t:\mathbf{z}_{t^-,i}|\mathbf{x}_{t-1})~,
\end{equation}
\noindent with $\mathbf{V}_{X_{t}} \backslash \mathbf{X}_{t-1} = \{ \mathbf{Z}_{t^-,1},\ldots,\mathbf{Z}_{t^-,G}\}$ indicating the set of $G$ past state variables of all processes $\mathbf{Z}_{t^-,i}$ that transfer information into the target variable $X_t$; note that $\mathbf{X}_{t-1}$, the history of the target, is explicitly not part of the set. The index $t^-$ is a reminder that only \textit{past} state variables are taken into account, i.e. $t^-<t$. As shown above, the local measures entering the sum are negative if they are mis-informative about the future of the target. Eventually the overall sum, or separable information, might also be negative, indicating that neither the pairwise information transfers, nor the history could explain the information contained in the target's future. This has been interpreted as a modification of either stored or transferred information. 

While this first attempt provided valuable insights in systems like elementary cellular automata \citep{Lizier2010a}, it is ultimately heuristic. A more rigorous approach is to look at decomposition of the local information $h(x_t)$ in the realization of a random variable to shed some more light on the issue which part of this information may be due to modification.
In this view, the overall information $H(X_t)$, in the future of the target process (or its local form, $h(x_t)$) can be explained by looking at \emph{all} sources of information and the history of the target \emph{jointly}, at least up to the genuinely stochastic part (innovation) in the target, as shown by \citet{Lizier2010a} (also see equations \ref{eq:EntropyDecomp}, \ref{eq:EntropyDecompLocal}). In contrast, we cannot decompose this information into \emph{pairwise} mutual information terms only. As described in the following, the remainder after exhausting pairwise terms is due to synergistic information between information sources and has motivated the suggestion to define information modification based on synergy \citep{Lizier2013InfoMod}.

To see the differences between a decomposition considering variables jointly or only in pairwise terms, consider a series of subsets formed from the set of all variables $\mathbf{Z}_{t^-,i}$ (defined above; ordered by $i$ here) that can transfer information into the target, except variables from the target's own history. The bold typeface in $\mathbf{Z}_{t^-,i}$ is a reminder that we work with a state space representation where necessary. Following the derivation by \citet{Lizier2010a}, we create a series of subsets 
$\mathbf{V}^g_{X_t}\backslash \mathbf{X}_{t-1}$ such that $\mathbf{V}^g_{X_t}\backslash \mathbf{X}_{t-1} =\{\mathbf{Z}_{t^-,1},\ldots,\mathbf{Z}_{t^-,g-1}\}$, i.e. the $g$-th subset only contains the first $g-1$ sources. We can decompose the collective transfer entropy from all our source variables, $TE(\mathbf{V}_{X_{t}}\backslash \mathbf{X}_{t-1} \rightarrow X_t)$, as a series of conditional mutual information terms, incrementally increasing the set that we condition on:
\begin{equation}
TE(\mathbf{V}_{X_{t}}\backslash \mathbf{X}_{t-1} \rightarrow X_t)=\sum_{g=1}^{G} I(X_t:\mathbf {Z}_{t^-,g}|\mathbf{X}_{t-1},\mathbf{V}^g_{X_t}\backslash \mathbf{X}_{t-1})~.
\end{equation}
The total entropy of the target $H(X_t)$ can then be written as:
\begin{equation}
\label{eq:EntropyDecomp}
 H(X_t)=A_{X_{t-1}}+\sum_{g=1}^{G} I(X_t:\mathbf{Z}_{t^-,g}|\mathbf{X}_{t-1},\mathbf{V}^g_{X_{t}}\backslash \mathbf{X}_{t-1})+W_{X_t}
\end{equation}
\noindent where $W_{X_t}$ is the genuine innovation in $X_t$. If we rewrite the decomposition in equation \ref{eq:EntropyDecomp} in its local form:
\begin{equation}
\label{eq:EntropyDecompLocal}
 h(x_t)=a_{X_{t-1}}+\sum_{g=1}^{G} i(x_t:\mathbf{z}_{t^-,g}|\mathbf{x}_{t-1},\mathbf{v}^g_{X_{t}}\backslash \mathbf{x}_{t-1})+w_{X_t}~,
\end{equation}
\noindent and compare to equation \ref{eq:sepInfo}, we see that the difference between the potentially mis-informative sum $s_{X_t}$ in equation \ref{eq:sepInfo} and the  fully accounted for information in  $h(x_t)$ from equation \ref{eq:EntropyDecompLocal} lies in the conditioning of the local transfer entropies. This means that the context that the source variables provide for each other is neglected and synergies and redundancies (see 
\secRef{PID}) 
are not properly accounted for. Importantly, the results of both equations (\ref{eq:sepInfo}, \ref{eq:EntropyDecompLocal}) are identical, if no information is provided either redundantly or synergistically by the sources $\mathbf{Z}_{t^-,g}$. This observation led Lizier and colleagues to propose a more rigorously defined measure of information modification based on the \emph{synergistic} part of the information transfer from the source variables $\mathbf{Z}_{t^-,g}$, and the targets history $\mathbf{X}_{t-1}$ to the target $X_t$ \citep{Lizier2013InfoMod}. This definition of information modification has several highly desirable properties. However, it relies on a suitable definition of synergy, which is currently only available for the case of two source variables (see \secRef{PID}). As there is currently a considerable debate on how to define the part of a the mutual information $I(Y:\{X_1,\ldots,X_i,\ldots\})$, that is synergistically provided by a larger set of source variables $X_i$, the question of how to best measure information modification maybe be considered open.

\section{Application examples}
\label{sec:AppEx}

\subsection{Active information storage in neural data}
\begin{figure}[bh]
\begin{center}
\includegraphics[width=11cm]{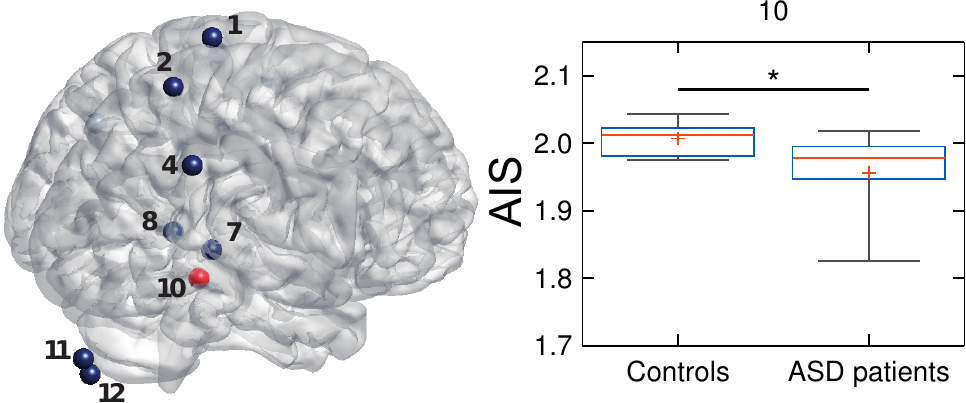}
\end{center}
\textbf{\refstepcounter{figure}\label{fig:ASD} Figure \arabic{figure}.}{ AIS in ASD patients compared to controls. (Left) Investigated MEG source locations (spheres; red = significantly lower AIS in ASD, blue = not sign.). (Right) Box and whisker plot for LAIS in source 10 (Hippocampus, corresponding to red sphere), where significant differences in AIS between patients and controls were found. \textit{Modified from \citep{Gomez2014ASDAIS}; creative commons attribution license (BB CY 3.0).} }
\end{figure}

Here, we present two very recent applications of (L)AIS to neural data and their estimation strategies for the PDFs. In both, estimation of (L)AIS was done using the JAVA information dynamics toolkit \citep{Lizier2014b,LizierJIDT} and state space reconstruction was performed in TRENTOOL \citep{Lindner2011} (for details, see \citep{Gomez2014ASDAIS,Wibral2014VSDLAIS}). The first study investigated AIS in magnetoencephalographic (MEG) source signals from patients with autism spectrum disorder (ASD), and reported a reduction of AIS in the hippocampus in patients compared to healthy controls \citep{Gomez2014ASDAIS} (Fig. \ref{fig:ASD}). In this study, the strategy for obtaining an estimate of the PDF was to use only baseline data (between stimulus presentations) to guarantee stationarity of the data. Results from this study align well with predictive coding theories \citep{Rao1999,Friston2006a} of ASD (also see \citet{Gomez2014ASDAIS}, and references therein). The significance of this study in the current context lies in the fact, that it explictely sought to measure the \emph{information processing consequences} at the algorithmic level of changes in neural dynamics in ASD at the implementation level.

\begin{figure}
\begin{center}
\includegraphics[width=12cm]{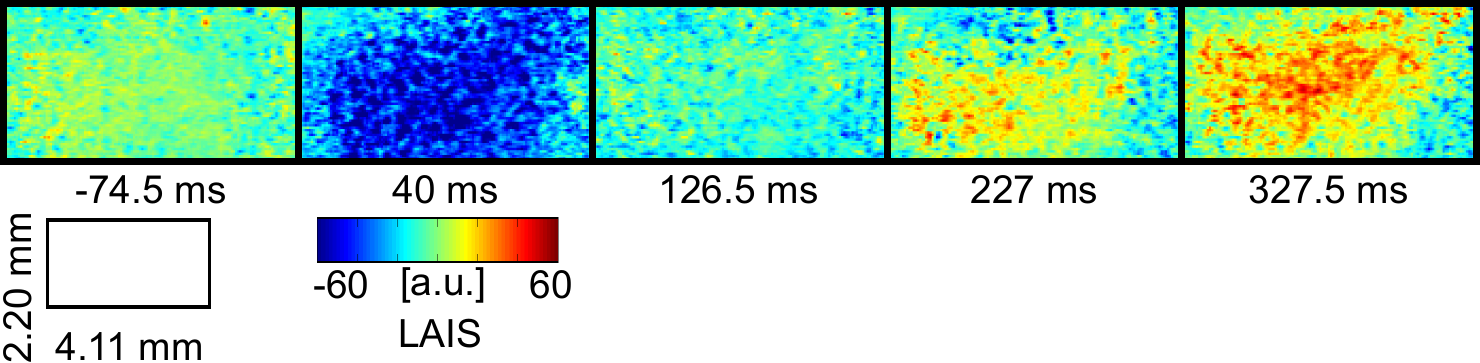}
\end{center}
\textbf{\refstepcounter{figure}\label{fig:VSD} Figure \arabic{figure}.}{ LAIS in VSD data from cat visual cortex (area 18), before and after presentation of a visual stimulus at time t=0ms. \textit{Modified from \citep{Wibral2014VSDLAIS}; creative commons attribution license (BB CY 3.0).}}
\end{figure}

The second study \citep{Wibral2014VSDLAIS} analyzed LAIS in voltage sensitive dye (VSD) imaging data from cat visual cortex. The study found low LAIS in the baseline before the onset of a visual stimulus, negative LAIS directly after stimulus onset and sustained increases in LAIS for the whole stimulation period, despite changing raw signal amplitude (Fig. \ref{fig:VSD}). In this study all available data were pooled, both from  baseline as well as stimulation periods, and also across all recording sites (VSD image pixels). Pooling across time is unusual, but reasonable insofar as neurons themselves also have to deal with nonstationarities as they arise, and a measure of \emph{neurally accessible} LAIS should reflect this. Pooling across all sites in this study was motivated by the argument that all neural pools seen by VSD pixels are capable of the same dynamic transitions as they were all in the same brain area. Thus, pixels were treated as physical replications for the estimation of the PDF. In sum, the evaluation strategy of this study is applicable to nonstationary data, but delivers results that strongly depend on the data included. Its future application therefore needs to be informed by precise estimates of the time scales at which neurons may sample their input statistics.

\subsection{Active Information Storage in a Robotic System}
\label{sec:AISRobotic}

Recurrent neural networks (RNNs) consist of a reservoir of nodes or artificial neurons connected in some recurrent network structure \citep{Maass2002,Jaeger2004}. Typically, this structure is constructed at random, with only the output neurons connections trained to perform a given task.
This approach is becoming increasingly popular for non-linear time-series modeling and robotic applications \citep{Dasgupta2013,Boedecker2012}.
The use of Intrinsic Plasticity based techniques \citep{Schrauwen2008} is known to assist performance of such RNNs in general, although this method is still outperformed on memory capacity tasks for example by the implementation of certain changes to the network structure \citep{Boedecker2009}.

To address this issue, \citet{Dasgupta2013} add an on-line rule to adapt the ``leak-rate'' of each neuron based on the AIS of its internal state. The leak-rate is reduced where the AIS is below a certain threshold, and increased where it is above.
The technique was shown to improve performance on delayed memory tasks, both for benchmark tests and in embodied wheeled and hexapod robots.
\citet{Dasgupta2013} describe the effect of their technique as speeding up or slowing down the dynamics of the reservoir based on the time-scale(s) of the input signal.
In terms of Marr's levels, we can also view this as an intervention at the algorithmic level, directly adjusting the level of information storage in the system in order to affect the higher-level computational goal of enhanced performance on memory capacity tasks.
It is particularly interesting to note the connection in information storage features across these different levels here.

\subsection{Balance of information processing capabilities near criticality}

It has been conjectured that the brain may operate in a self-organized critical state \citep{Beggs2003}, and recent evidence demonstrates that the human brain is at least very close to criticality, albeit slightly sub-critical \citep{Priesemann2013,Priesemann_spike_2014}.
This prompts the question of what advantages would be delivered by operating in such a critical state.
From a dynamical systems perspective, one may suggest that the balance of stability (from ordered dynamics) with perturbation spreading (from chaotic dynamics) in this regime \citep{Langton1990} gives rise to the scale-free correlations and emergent structures that we associate with computation in natural systems.
From an information dynamics perspective, one may suggest that the critical regime represents a balance between capabilities of information storage and information transfer in the system, with too much of either one decaying the ability for emergent structures to carry out the complementary function \citep{Lizier2011,Lizier2008RBNs,Langton1990}.

Several studies have upheld this interpretation of maximised but balanced information processing properties near the critical regime. 
In a study of random Boolean networks it was shown that TE and AIS are in an optimal balance near the critical point \citep{Lizier2008RBNs,Lizier2011}. This is echoed by findings for recurrent neural networks \citep{Boedecker2012} and for maximisation of transfer entropy in the Ising model \citep{Barnett2013}, and maximization of entropy in neural models and recordings \citep{Haldeman2005,Shew2013}.
From Marr's perspective, we see here that at the algorithmic level the optimal balance of these information processing operations yields the emergent and scale-free structures associated with the critical regime at the implementation level.
This reflects the ties between Marr's levels as described in \secRef{AISRobotic}.
These theoretical findings on computational properties at the critical point are of great relevance to neuroscience, due to the aforementioned importance of criticality in this field.

\subsection{Local information dynamics in Cellular Automata}
\label{sec:CAs}

Cellular automata (CAs) are discrete dynamical systems with an array of cells
that synchronously update their value as a function of a fixed number of spatial
neighbours cells, using a uniform rule \citep{WolframBook}.
CAs are a classic complex system where, despite their simplicity, emergent structures arise.
These include \emph{gliders}, which are coherent structures moving against regular background domains.
These gliders and their interactions have formed the basis of analysis of cellular automata as canonical examples of nature-inspired distributed information processing (e.g. in a distributed ``density'' classification process to determine whether the initial state had a majority of ``1'' or ``0'' states) \citep{Mitchell1998}.
In particular, (moving) gliders were conjectured to transmit information across the CA, static gliders to store information, and their collisions or interactions to process information in ``computing'' new macro-scale dynamics of the CA.

Local transfer entropy, active information storage and separable information were applied to CAs to produce spatiotemporal local information dynamics profiles in a series of experiments \citep{Lizier2008,Lizier2010a,Lizier2012a,Lizier2014,LizierBook}. The results of these experiments confirmed the long-held conjectures that gliders are the dominant information transfer entities in CAs, while blinkers and background domains are the dominant information storage components, and glider/particle collisions are the dominant information modification events.
These results are crucial in demonstrating the alignment between our qualitative understanding of emergent information processing in complex systems and our new ability to quantify such information processing via these measures.
These insights could only be gained by using local information measures, as studying averages alone tells us nothing about the presence of these spatiotemporal structures.

\newcommand{\caWidth}{0.23\textwidth}
\newcommand{\caHeight}{0.17\textwidth}
\begin{figure*}
  \begin{center}
		\subfigure[Raw CA]{\label{fig:phi-raw}\makebox[\caWidth]{\includegraphics[trim= 0 0 120 0,clip=true,height=\caHeight]{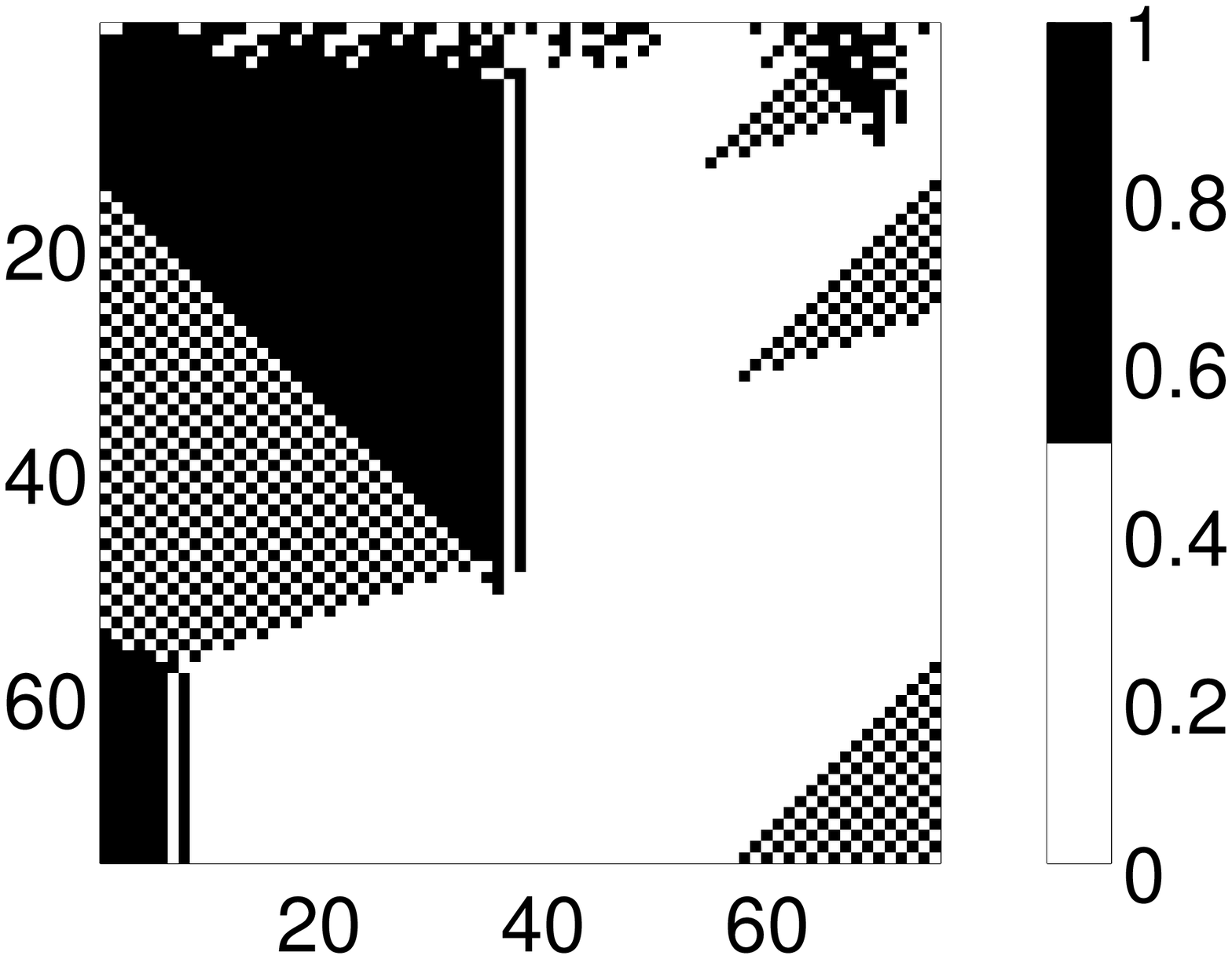}\ \ \ \ \ \ \ \ }} \ \ 
		\subfigure[$a(i,n,k=10)$]{\label{fig:phi-active-colour-10}\makebox[\caWidth]{\includegraphics[height=\caHeight]{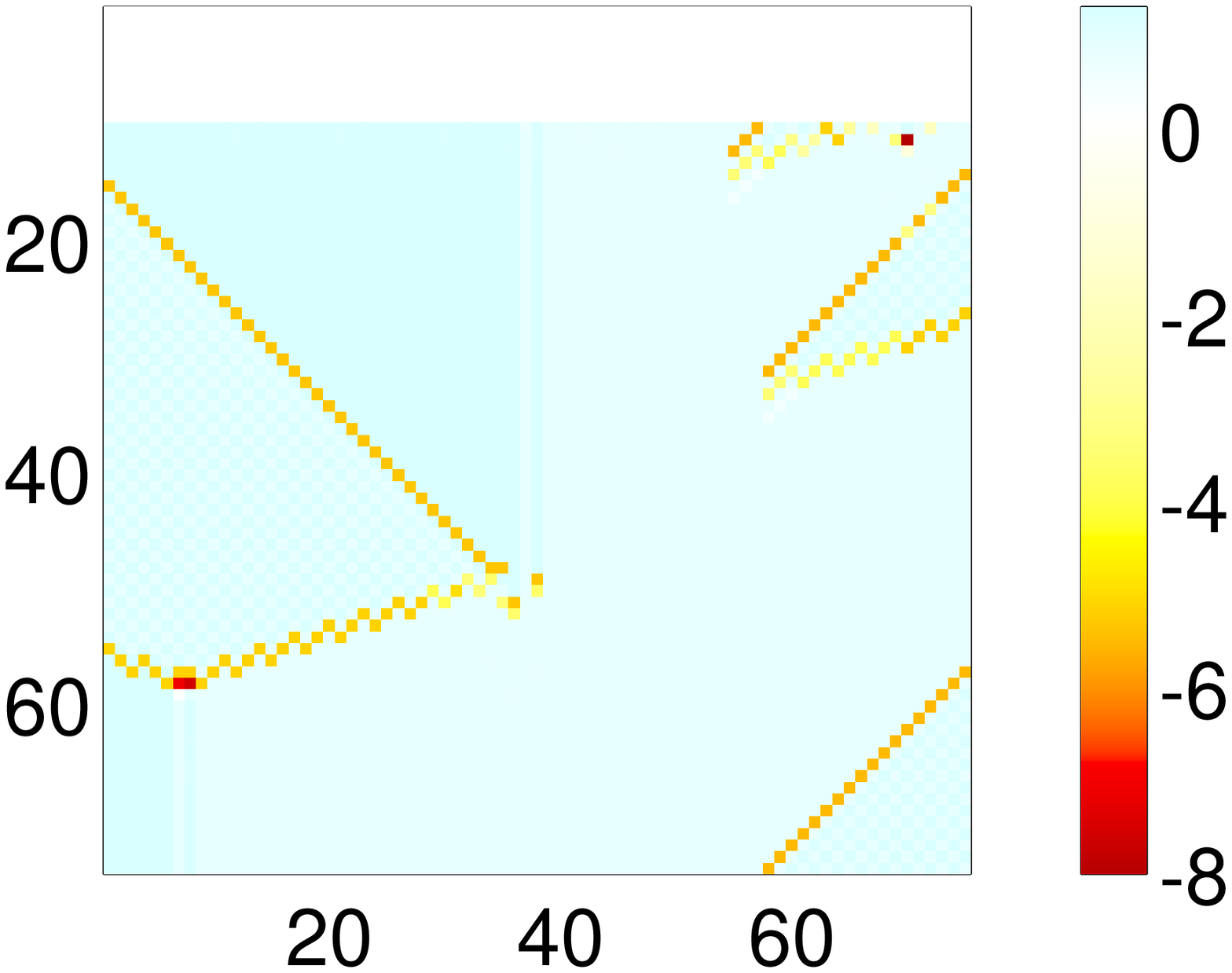}}}
		
		\subfigure[$t(i,j=-1,n,k=10)$]{\label{fig:phi-te--1-colour-10}\makebox[\caWidth]{\includegraphics[height=\caHeight]{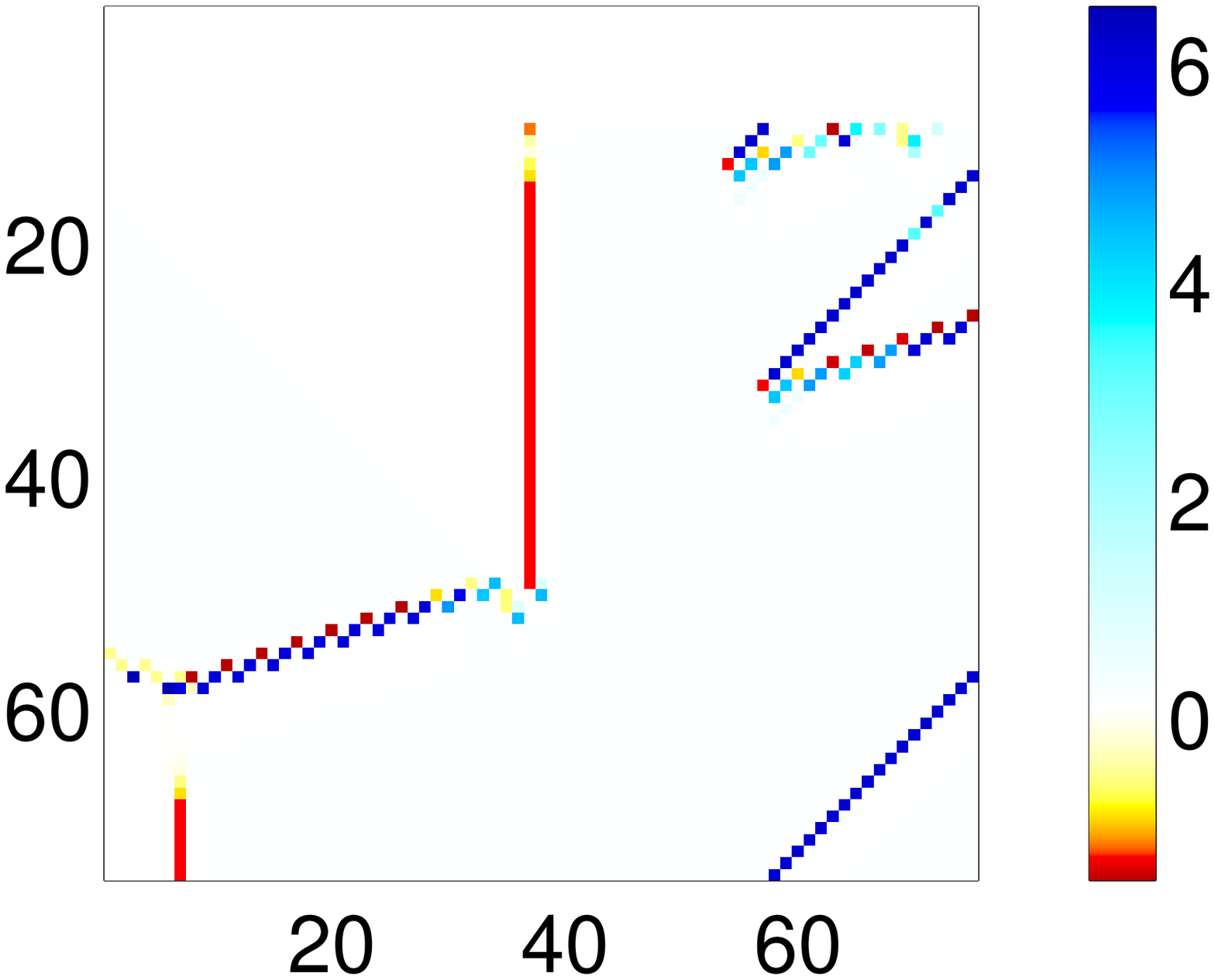}}} \ \ 
		\subfigure[$t^c(i,j=-1,n,k=10)$]{\label{fig:phi-teComp--1-colour-10}\makebox[\caWidth]{\includegraphics[height=\caHeight]{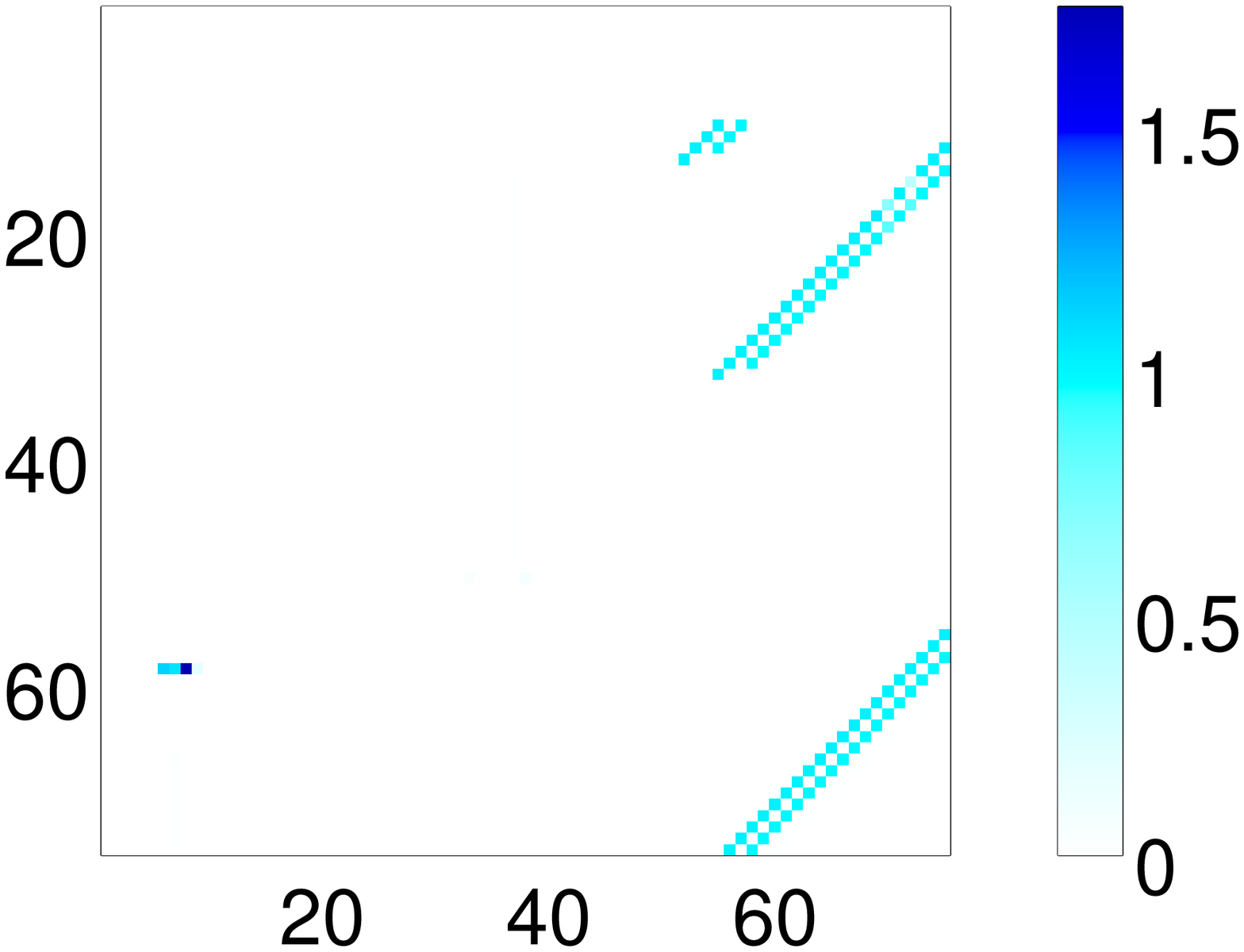}}}
		
	\end{center}
	\textbf{\refstepcounter{figure}\label{fig:phi} Figure \arabic{figure}.}{ Local information dynamics in rule $\mathbf{\phi_{par}}$. Local information dynamics in $r=3$ rule $\phi_{par}$ for the raw values displayed in \subref{fig:phi-raw} ~~(black for ``1'', white for ``0''). 75 time steps are displayed for 75 cells, starting from an initial random state. Notice that a short initial transient occurs after which the emergent structures arise. 
	For the spatiotemporal information dynamics plots (\subref{fig:phi-active-colour-10}~--\subref{fig:phi-teComp--1-colour-10}~), we use a history length $k=10$ (therefore the measures are undefined and not plotted for $n \leq 10$), and all units are in bits.
	We have:
		\subref{fig:phi-active-colour-10} ~~Local active information storage $a(i,n,k=10)$; 
		\subref{fig:phi-te--1-colour-10} ~~Local apparent transfer entropy one cell to the left $t(i,j=-1,n,k=10)$;
		and
		\subref{fig:phi-teComp--1-colour-10} ~~Local complete transfer entropy one cell to the left $t^c(i,j=-1,n,k=10)$.
		(After \citet{Lizier2014c}.)
	}
\end{figure*}

For our purposes, a crucial step was the extension of this analysis to a CA rule (known as $\phi_{par}$) which was evolved to perform the density classification task outlined above \citep{LizierBook,Lizier2014c}, since we may interpret this with Marr's levels (\secRef{UseInfoNeuro}).
Spatiotemporal profiles of local information dynamics for a sample run of this density classification rule are shown in \fig{phi}, and may be reproduced using the \texttt{DemoFrontiersBitsFromBiology2014.m} script in the \texttt{demos/\-octave/\-CellularAutomata} demonstration distributed with the Java Information Dynamics Toolkit \citep{Lizier2014b}.
In this example, the classification of the density of the initial CA state is the clear goal of the computation (task level).  At the algorithmic level, our local information dynamics analysis allowed direct identification of the roles of the emergent structures arising on the CA  after a short initial transient \fig{phi}. For example, this analysis revealed markers that CA regions had identified local majorities of ``0'' or ``1'' (see the wholly white or black regions, or checkerboard patterns indicating uncertainty). These regions are identified as storing this information in \fig{phi-active-colour-10}.
The analysis also quantifies the role of several glider types in communicating the presence of these local majorities and the strength of those majorities (see the slow and faster glider structures identified as information transfer in \fig{phi-te--1-colour-10} and \fig{phi-teComp--1-colour-10}), and the role of glider collisions resolving competing local majorities.

\subsection{Information cascades in swarms and flocks}
\label{sec:Swarms}

Swarming or flocking refers to the collective behaviour exhibited in movement by a group of animals \citep{Lissaman1970,Parrish1999}, including the emergence of patterns and structures such as cascades of perturbations travelling in a wave-like manner, splitting and reforming of groups and group avoidance of obstacles.
Such behaviour is thought to provide biological advantages, e.g. protection from predators.
Realistic simulation of swarm behaviour can be generated using three simple rules for individuals in the swarm, based on separation, alignment and cohesion with others \citep{Reynolds1987}.

\citet{Wang2012} analysed the local information storage and transfer dynamics exhibited in the patterns of motion in a swarm model, based on time-series of (relative) headings and speeds of each individual.
Most importantly, this analysis quantitatively revealed the coherent cascades of motion in the swarm as waves of large, coherent information transfer (as had previously been conjectured, e.g. see \citet{Couzin2006} and \citet{Bikhchandani1992}).

These ``information cascades'' are analagous to the gliders in CAs (above).
When viewed using Marr's levels they have a similar algorithmic role of carrying information coherently and efficiently across the swarm, while the implementation of the information here is simply in the relative heading and speed of the individuals.
The goal of the computation
(task level) for the swarm depends on the current environment, but may be to avoid predators, or efficiently transport the whole group to nesting or food sites.

\subsection{Transfer entropy guiding self-organisation in a Snakebot}
\label{sec:Snakebot}

\citet{Lizier2008Snake} inverted the usual use of transfer entropy, applying it for the first time as a \emph{fitness function} in the evolution of adaptive behaviour, as an example of \emph{guided self-organisation} \citep{Prokopenko2009b,Prokopenko2014}.
This experiment utilised a \emph{snakebot} -- a snake-like robot with separately controlled modules along its body, whose individual actuation was evolved via genetic programming (GP) to maximise transfer entropy between adjacent modules.
The actual motion of the snake emerged from the interaction between the modules and their environment.
While the approach did not result in a particularly fast-moving snake (as had been hypothesised), it did result in coherent travelling information waves along the
snake, which were revealed only by local transfer entropy.

These coherent information waves are akin to gliders in CAs and cascades in swarms (above), suggesting that such waves may emerge as a resonant mode in evolution for information flow.
This may be because they are robust and optimal for coherent
communication over long distances, and may be simple to construct via
evolutionary steps.
Again, we may use Marr's levels here to identify the goal of the computation (task level) as to transfer information between the snake's modules here (perhaps information about the terrain encountered).
At the algorithmic level the coherent waves carry this information efficiently along the snake's whole body, while the implementation is simply in the attempted actuation of the modules on joints and their interaction (tempered by the environment).

\section{Conclusion and Outlook}
Neural systems perform acts of information processing in the form of distributed (biological) computation, and many of the more complex computations and emergent information processing capabilities remain mysterious to date. Information theory can help to advance our understanding in two ways. On the one hand, neural information processing can be decomposed into its component processes of information storage, transfer and modification using information theoretic tools. This allows us to derive constraints on possible algorithms served by the observed neural dynamics.  On the other hand, the representations that these algorithms operate on, can be guessed by analyzing the mutual information between human-understandable descriptions of relevant concepts and quantities in our experiments and indices of neural activity. This helps to identify which parts of the real world neural systems care for. However, care must be taken when asking such questions about neural codes as the question of how neurons code jointly has not been solved completely to date. Taken together, the knowledge about representations and possible algorithms describes the operational principles of neural systems at Marr's algorithmic level and may hint at solutions for solving ill-defined real world problems that biologically inspired computing systems have to face with their constrained resources.

\section*{Disclosure/Conflict-of-Interest Statement}
The authors declare that the research was conducted in the absence of any commercial or financial relationships that could be construed as a potential conflict of interest.

\section*{Author Contributions}
VP, MW and JL wrote and critically revised the manuscript.

\section*{Acknowledgement}
We thank Patricia Wollstadt and Lucas Rudolf for proof reading of the manuscript.
\paragraph{Funding\textcolon} MW was supported by LOEWE-Grant ``Neuronale Koordination Forschungsschwerpunkt Frankfurt''. VP received financial support from the German Ministry for Education and Research (BMBF) via the Bernstein Center for Computational Neuroscience (BCCN) G\"{o}ttingen
under Grant No. 01GQ1005B. 

\bibliographystyle{frontiersinSCNS&ENG} 

\end{document}